\documentclass[twocolumn]{aa}  

\usepackage{graphicx}
\usepackage{txfonts}
\usepackage{epstopdf}
\usepackage{xspace} 
\usepackage{natbib}
\usepackage{mathrsfs}

\newcommand{\Msun}{$M_\odot$}

\newcommand{\herschel}{\textit{Herschel}\xspace}

\newcommand{\planck}{\textit{Planck}\xspace}

\newcommand{\micron}{\,$\mu$m\xspace}

\begin{document}

\title{Fourier-space combination of \planck and \herschel images}

\author{J. Abreu-Vicente\inst{1}\thanks{Member of the
International Max Planck Research School (IMPRS) at the University
of Heidelberg and the SFB 881 project, ``The Milky Way System''}, A. Stutz\inst{1,2}, Th. Henning\inst{1},
E. Keto\inst{3}, J. Ballesteros--Paredes\inst{4,5} \& T. Robitaille\inst{1}}

\authorrunning{J. Abreu-Vicente, A. Stutz et al.}
\titlerunning{Multi-scale \planck corrections to \herschel}

\offprints{J. Abreu-Vicente, \email{abreu@mpia-hd.mpg.de}}

\institute{$^{1}$Max-Planck-Institut f\"{u}r Astronomie (MPIA),
K\"{o}nigstuhl 17, 69117, Heidelberg, Germany\\
$^{2}$Departamento de Astronom\'{i}a, Universidad
de Concepci\'{o}n, Av. Esteban Iturra s/n, Distrito Universitario, 160-C, Chile\\
$^{3}$Harvard-Smithsonian Center for Astrophysics, Cambridge, MA, 02138, USA\\
$^{4}$ Instituto de Radioastronom\'{i}a y Astrof\'{i}sica, Universidad Nacional Aut\'{o}noma de M\'{e}xico, Campus Morelia Apartado Postal 3-72, 58090, Morelia, Michoac\'{a}n, M\'{e}xico\\
$^{5}$Universit\"{a}t Heidelberg, Zentrum f\"{u}r Astronomie, Institut f\"{u}r Theoretische Astrophysik, Albert-Ueberle-Str. 2, D-69120 Heidelberg, Germany}

\date{Received ; accepted }

\abstract{\herschel has revolutionized our ability to measure column densities
(N$_{\rm H}$) and temperatures (T) of molecular clouds thanks to its
far infrared multiwavelength coverage. However, the lack of a well
defined background intensity level in the Herschel data limits the
accuracy of the N$_{\rm H}$ and T maps.  }
{We aim to provide a method that corrects the missing \herschel
  background intensity levels using the \planck model for foreground Galactic thermal
  dust emission.  For the \herschel/PACS data, both the
  constant--offset as well as the spatial dependence of the missing
  background must be addressed.  For the \herschel/SPIRE data, the
  constant--offset correction has already been applied to the archival
  data so we are primarily concerned with the spatial dependence, which
  is most important at 250~$\mu$m.}  {We present a Fourier method
  that combines the publicly available \planck model on large angular
  scales with the \herschel images on smaller angular scales.}  {We apply our method to two regions spanning a
  range of Galactic environments: Perseus and the Galactic plane
  region around $l = 11\deg$ (HiGal--11).  We post-process the
  combined dust continuum emission images to generate column density
  and temperature maps. We compare these to previously adopted
  constant--offset corrections. We find significant differences
  ($\gtrsim$20\%) over significant ($\sim$15\%) areas of the maps, at
  low column densities ($N_{\rm H}\lesssim10^{22}$\,cm$^{-2}$) and
  relatively high temperatures ($T\gtrsim20$\,K). We also apply our method to
  synthetic observations of a simulated molecular cloud to validate
  our method.}  {Our method successfully corrects the \herschel
  images, including both the constant--offset intensity level and the
  scale-dependent background variations measured by \planck. Our
  method improves the previous constant--offset corrections, which did
  not account for variations in the background emission levels.}


\maketitle

\section{Introduction}

The {\it Herschel} Space Telescope PACS~\citep{pog10} and
SPIRE~\citep{griffin10} photometers have surveyed large areas of the
sky~\citep[e.g.,
][]{andre10,gordon10,kramer10,meixner10,molinari10,fritz12,draine14,stutz16}
in the far-infrared (FIR) and sub-millimeter (sub--mm) from 70 
to 500~\micron, measuring the cold dust emission largely inaccessible
from the ground.  Furthermore, the stability of space-based
observations allows for the recovery of extended emission down to much
fainter flux levels and over larger scales than those accessible with
ground-based sub-mm data.  Simultaneously, the \herschel data probe
higher column densities at higher resolution than those commonly
accessible with near-infrared (NIR) extinction measurements~\citep[but
see also][]{stutz09,kainul11}.

However, even given the wealth of information that the \herschel PACS
and SPIRE continuum data provide, large portions of these data remain
to be fully scientifically exploited. One obstacle to obtaining
accurate column density and temperature maps is that the \herschel
archive data have not received a full background correction.
Obtaining such corrections is not trivial.  In the case of the SPIRE
images, the archive data have been partially corrected with a
\planck--derived~\citep{2014A&A...571A...1P}
constant--offset\footnote{This procedure is described in detail in the
  instrument handbook:
  \texttt{herschel.esac.esa.int/Docs/SPIRE/spire\_
    handbook.pdf.}}. The constant--offset correction for SPIRE assumes
average zero--level flux values (a single constant-offset correction
over a given map), based on \planck measurements.  Similar corrections
have also been applied in~\citet{bernard10} and
\citet{lombardi14}. Furthermore, we also refer to~\citet{zari16} for a
near infrared extinction and \planck based calibration method. All
these methods implicitly assume that the corrections to the \herschel
intensity are independent of angular scale.  We note that comparisons
with \planck data show that this is in general a good approach for the
SPIRE 350 and 500~\micron SPIRE data~\citep[][]{bertincourt16}.  In
the case of the \herschel archive PACS images no background corrections
have been applied.

While the constant-offset corrections partially account for the
missing background in the \herschel images, the \planck flux
distribution may significantly vary within the image
area, especially in cases where the maps are large and at shorter
wavelengths.  To our knowledge there is no previous demonstration that
the constant--offset correction will fully capture the background
variations in the PACS and SPIRE 250~\micron data.  Therefore, both
PACS and SPIRE images would benefit from a background correction that
is capable of grasping the scale dependence of the background emission
levels.  The knowledge of these background levels are an obvious 
requirement to estimate the
``actual'' flux scale measurements in \herschel images.

The \planck all-sky dust model~\citep{planckXI14} is currently the
best available option for correcting the \herschel images in the
wavelength range 160\micron--500~\micron because of the close match in
wavelength coverage. The dust model obtained from \planck was derived
using 353~GHz, 545~GHz, 857~GHz, and IRAS 100~\micron data.  The
inclusion of the IRAS 100~\micron data in the \planck model helps to
better constrain peak of the dust spectral energy distribution near
$\sim\,$160~\micron.  Here we develop a method that uses this \planck
model to correct the arbitrary flux scale of the \herschel data.  In
the case of PACS, this correction includes both constant--offset as
well as spatial dependence of the corrections that capture the
variations in the \planck fluxes at large scales.  In the case of
SPIRE, since these data already include the constant--offsets from
\planck, the correction addresses the possible spatial variations in
the background levels.  Thus, this method is specially relevant for
the PACS data, but can also be important for SPIRE, and in particular
for the 250\micron data.  In summary, here we essentially combine the
\planck and \herschel maps in Fourier space, keeping the information
of the former at large scales and the latter at small scales. The
transition from large (\planck) to small (\herschel) scales, defined
as scale at which \planck and \herschel have similar amplitudes in the
Fourier space, is individually estimated for each map.

We apply our method to two fields observed by \herschel that span a
wide range of Galactic environments: Perseus, and the Galactic plane
region at $l = 11\degr$ (HiGal--11, including G11 and W31).  We use
the new background-calibrated maps to obtain dust column density
(N$_{\rm H}$) and temperature (T) maps. We compare our column density
maps to those obtained from \herschel maps corrected with the 
constant--offset method alone. The data processed in this work are publicly
available.

This paper is organized as follows. We describe the
data used in Sect.~\ref{sec:data}. In Sect.~\ref{sec:method} we
describe our Fourier technique and its application to the \herschel
and \planck data. In Sect.~\ref{sec:results} we show the flux maps
obtained with our methodology and post--process these to obtain column
density (N$_{\rm H}$) and temperature (T) maps.  In
Sect.~\ref{sec:disc} we compare our results with previous methods, 
testing the performance of our method with simulated data in 
Sect.~\ref{sec:sims}.  We present our conclusions in Sect.~\ref{sec:conc}.

\begin{table*}[t]
  \caption{\herschel parallel mode observations analyzed in this paper.} 
  \centering 
  \begin{tabular}{c c c c c c c} 
    \hline\hline 
    Name			& Obs ID		& RA	 (J2000)		& DE	 (J2000)		& Map size & Project				& Ref\\
    &			& [hh:mm:ss]	& [$\degr$:$\arcmin$:$\arcsec$] & [$\arcmin\times\arcmin$]&  &\\\hline
    Perseus-04	& 1342190326& 03:29:39	& +30:54:34	& $138\times138$& KPGT\_ pandre\_ 1	&1 , 2\\
    HiGal-11    &1342218966& 18:09:50	& -19:25:22 & $72\times72$ & KPOT\_ smolinar\_ 1&5\\\hline
    \label{tab:obs}
  \end{tabular}
  \tablebib{(1)~\citet{andre10}; (2)~\citet{sadavoy13}; 
    (3)~\citet{motte10}; (4)~\citet{motte12}; (5)~\citet{molinari10}.} 
\end{table*}

\section{Data}\label{sec:data}

In this paper we use public \herschel \citep{pil10} and \planck
\citep{planckI13} archive data.  

\subsection{\herschel data}\label{data:herschel}

The \herschel data used in this paper were retrieved from the
\herschel science archive. We select parallel mode observations
carried out with the PACS~\citep{pog10} and SPIRE~\citep{griffin10}
photometers.  We use the level 2.5 data products.  These data products
are optimized for extended emission reconstruction as well as the
principle observing mode used for large-scale surveys (i.e., the
parallel mode).  We therefore focus exclusively on these products in
this paper. We use the red (160\,$\mu$m) channel of PACS, and the
three wavelengths of SPIRE (250\,$\mu$m, 350\,$\mu$m, and
500\,$\mu$m).  These maps have native pixel scales (and beam sizes) of
3.2$\arcsec$ (11.8$\arcsec$), 6$\arcsec$ (18.2$\arcsec$), 10$\arcsec$
(24.9$\arcsec$) and 14$\arcsec$ (36.3$\arcsec$) respectively.  We
refer the reader to Table~\ref{tab:obs} for further details.

\subsection{\planck all-sky foreground dust emission model}\label{data:planck}

The \planck satellite has observed the entire sky at nine different
frequencies in the range 30 -- 857\,GHz~\citep{planckI13}.  
Since \herschel and \planck instruments do not have similar wavelength 
coverage, we need to convert the \planck observations into
maps directly comparable to \herschel. One of the data products 
of the \planck mission is an all-sky model of the foreground dust
emission, obtained from a modified blackbody (MBB) fit to \planck
observations at 353, 545, and 857 GHz, complemented with IRAS
100\,$\mu$m~\citep{beichman88} observations~\citep{planckXI14,2016A&A...586A.132P}.  This
model estimates the dust optical depth, temperature, and spectral
index with a resolution of 5$\arcmin$ (30$\arcmin$ for the spectral
index, $\beta$). The results of this model should be used only within the
frequency range 353--3000\,GHz. At shorter wavelengths the dust emission
is known to contain a non-thermal component due to stochastically heated 
grains~\citep[e.g. ][]{draine07,draine11,planckXI14,meisner15}.

We use the \planck all-sky foreground dust emission model to
reconstruct a FIR spectral energy distribution (SED) at the observed
\herschel wavelengths.  This model provides the optical depth at
$\nu_{0}=353$\,GHz ($\tau_{0}$), the dust temperature
($T_{\mathrm{obs}}$), and the dust spectral index ($\beta$) for each
sky pixel based on a MBB fit to the observed
fluxes.  We obtain the SED following the \planck analysis via
\begin{equation}\label{eq:sed}
I_{\nu} = B_{\nu}(T_{\mathrm{obs}})\tau_{0} \left ( \frac{\nu}{\nu_{0}} \right )^{\beta}, 
\end{equation}
where $I_{\nu}$ is the intensity at each frequency, and
$B_{\nu}(T_{\mathrm{obs}})$ is the blackbody function at the observed
temperature.  We convert these SEDs into \herschel simulated
observations, integrating them over the respective \herschel filter
response functions for extended sources.  The \herschel pipeline
assumes a flat $\nu S_{\nu}$ calibration within each bandpass. We
therefore obtain the monochromatic \planck fluxes ($S$) as follows:
\begin{equation}\label{eq:int-sed}
S = \frac{\int {I_{\nu}\,R_{\nu}\,\mathrm{d}\nu}}{\int \frac{\nu_0}{\nu}R_{\nu}\,\mathrm{d}\nu},
\end{equation}
where $I_{\nu}$ is the intensity obtained in Eq.~\ref{eq:sed},
$R_{\nu}$ is the spectral response function for each \herschel
bandpass, and $\nu_{0}$ the effective central frequency of each
bandpass~\citep{robitaille07}. We repeat this step for each pixel of
the \planck all-sky dust emission model, obtaining four maps of
simulated emission at the targeted \herschel wavelengths. These maps
are initially extracted from the \planck healpix data format at a
75\arcsec pixel scale.  In a later step these images are regrided and
rotated to the reference frame of the \herschel images at their
respective wavelengths (pixel scales for \herschel data are listed
above).  For simplicity, we refer to this data cube as the \planck
data cube in the reminder of the paper.  

For completeness, we investigate how the uncertainties of the
parameters $T_{\mathrm{obs}}$, $\beta$, and $\tau_{0}$ propagate into
our simulated flux maps. To estimate the effect of uncertainties we
use the standard deviations of $T_{\mathrm{obs}}$, $\beta$, and
$\tau_{0}$ derived for the whole sky, which are respectively 8\%, 8\%,
and 10\% \citep{planckXI14}. We apply these values to a MBB function
independently and estimate how much the flux varies at the four
wavelengths of interest.  In Table~\ref{tab:unc} we show the results
for the representative fiducial MBB parameters
$T_{\mathrm{obs}} = 20$, $\beta = 1.7$, and $\tau_{0} = 10^{-4}$. At
every wavelength, the temperature uncertainties dominate on our
simulated maps, with the effects being larger at shorter
wavelengths. The flux errors caused by the dust spectral index and
optical depth uncertainties are within calibration errors of the
instruments.

\begin{figure}
\resizebox{\hsize}{!}{\includegraphics{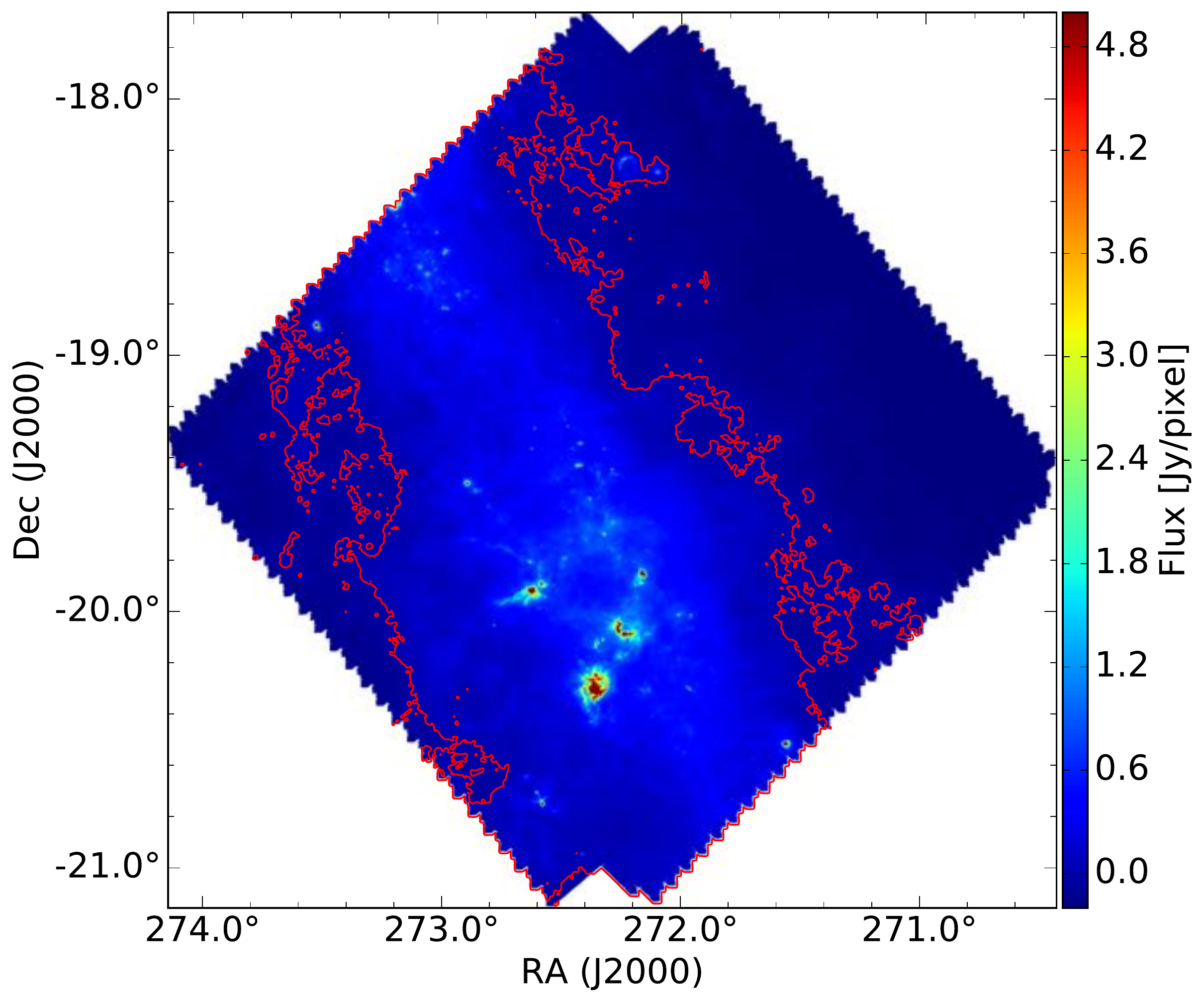}}
\caption{\herschel / PACS 160\,$\mu$m map of HiGal--11.  The
  red contour shows the region with negative fluxes
  in the \herschel map available in the \herschel data 
  archive. This image also shows the ``saw effect'' in the
  map borders and the zero--pading (white edges surrounding
  the map).}
  \label{fig:crop}
\end{figure}

\begin{table}
  \caption{Uncertainties in flux values as propagated from the \planck dust model.} 
  \centering 
  \begin{tabular}{c c c c} 
    \hline\hline 
    $\lambda$ [$\mu$m]	& $\sigma_{T}=8\%$	& $\sigma_{\beta}=8\%$	& $\sigma_{\tau}=10\%$\\\hline
    160					& 42\%				& 5\%					& 10\% 					\\
    250					& 29\%				& 4\%					& 10\%					\\
    350					& 22\%				& 4\%					& 10\%					\\
    500					& 15\%				& 3\%					& 10\%					\\\hline
    \label{tab:unc}
  \end{tabular}\\
  \footnotesize{Uncertainties in the model fluxes for fiducial MBB 
    parameters of $T_{\mathrm{obs}} = 20$, $\beta = 1.7$, and $\tau_{0} = 1e-4$. }
\end{table}

\subsection{Initial image processing}\label{data:init}

In the main step of this method (see below Sect.~\ref{sec:method}) we
combine the \planck and \herschel datasets obtained above in Fourier
space (see Sect.~\ref{sec:comb}). Fourier Transforms (FTs) are sensitive
to any spatial patterns in the maps. As we show in
Fig.~\ref{fig:crop}, the original \herschel maps have two main spatial
patterns: a ``saw'' effect in the field edges, and a zero--padding
outside of the observed region.  Before applying our method, we rotate
and crop the \herschel maps in order to avoid possible contamination
of the FT amplitudes from the zero padding and the saw edges.
Unfortunately, the general field geometry of \herschel data is not
well described by a rectangular field.  We therefore must find the
best combination possible between removing zero-padding and saw
effects and keeping the largest image area as possible. The \herschel
/SPIRE and \herschel /PACS observations in parallel mode have an
intrinsic pointing offset\footnote{ See \herschel handbook for further
  details.}.  We therefore treat both instruments separately and
define different effective regions for each instrument. The following
steps are applied to the image products of this initial processing.
We address possible effects of these initial processing steps in
Sect.~\ref{sec:sims}.

\section{Method and implementation}\label{sec:method}

The main goal of this paper is to derive \planck--based multi--scale
corrections for the \herschel images at each observed wavelength.
Here we combine two single-dish datasets: one with much better
resolution (\herschel) than the other (\planck).  Therefore, we can
make a loose analogy to previous interferometric techniques
\citep[e.g.,][]{thompson86,stanimirovic02} aimed at combining data
sets that are observed at significantly different angular resolutions.
Here \herschel would represent the interferometer data while \planck
would represent the lower resolution single-dish data.  Furthermore,
our method is similar to previous methods combining single dish
observations, e.g., \planck and ATLASGAL \citep{csengeri16}.

\subsection{Cross--calibration and combination of the datasets}\label{sec:crosscal}

Before combining images it is crucial that both have the same or
similar flux scales. This step ensures no  sharp jumps in intensities that
may cause artifacts when combining the data in Fourier space.
Due to the linearity of the Fourier transform, the cross--calibration
can be done either in the image-- or in the \emph{uv}--plane,
where both methods are mathematically equivalent.  The procedure to
cross--calibrate \planck and \herschel in the image plane consists of 
applying a linear fit $y = mx + b$ to the \herschel and \planck
datasets and apply the constant $b$ to the \herschel data
\citep[e.g.,][]{bernard10,lombardi14,zari16}.  This is equivalent to
correcting the \herschel dataset using only the zeroth Fourier mode.
As above, we refer to this image plane correction as the
``constant--offset'' technique. If a constant--offset would be the only
difference between both datasets, this correction would be
sufficient. This method has already been applied to the SPIRE data
products in the \herschel Science Archive (see above).

Alternatively, here we take advantage of the overlap of the \herschel
and \planck datasets in Fourier space and cross--calibrate the data by
comparing their relative Fourier amplitudes in the \emph{uv}--plane.
First, we re--grid the \planck data (75$\arcsec$/pixel) to the
corresponding \herschel pixel scale.  We then Fourier transform both
datasets. To be able to compare them, we must convolve the \planck
visibilities with the \herschel beam. This convolution is achieved in
two steps: \emph{i)} by first deconvolving the \planck visibilities
i.e., dividing them by the \planck beam, assumed to have a
FWHM\,$=5\arcmin$; \emph{ii)} convolving (multiplying) the resulting
visibilities with the corresponding \herschel beam profile. Note that
the convolution of the \planck visibilities implies dividing by an
exponential function that approaches zero at small scales,
exponentially increasing the noise of the \planck data at small
scales. However, we are only interested in the large scales where the
noise is not significantly amplified by the beam deconvolution.

\begin{figure}
\resizebox{\hsize}{!}{\includegraphics{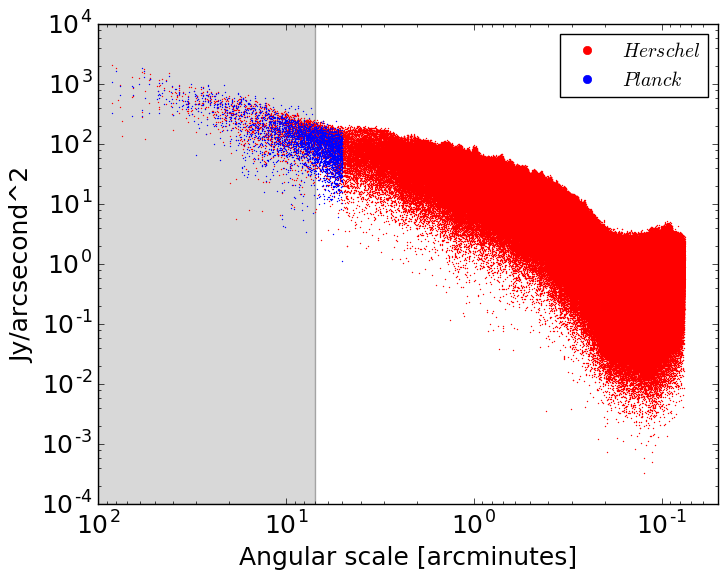}}
\caption{\herschel (red) and deconvolved \planck (blue) visibilities of the
	field HiGal--11 at 160\,$\mu$m. The fluxes are
	in units of Jy/arcsec$^{2}$ and the scales in
	units of arcminutes, both shown in logarithmic 
	scale. The shadowed region between 7$\arcmin$ and
	100$\arcmin$ show the visibilities used to obtain
	the cross--calibration factor between both datasets. 
	We only show \planck visibilities at scales
	larger than 5$\arcmin$ to avoid the noise increment
	at smaller scales caused by the deconvolution of the
	\planck data. }
  \label{fig:crossCal}
\end{figure}

The \herschel and \planck visibilities are shown in
Figure~\ref{fig:crossCal}. We compare their visibilities at scales on
which the signal--to--noise ratio of both datasets is high enough to
obtain the calibration factor to be applied. The high
signal--to--noise ratio requirement limits us to compare the
visibilities at scales larger than the 5$\arcmin$ resolution of
\planck. In order to be conservative and avoid the noise contamination
generated by the deconvolution of the \planck data, we will define the
smallest scale at which we compare the \planck at \herschel
visibilities at 7$\arcmin$.  At these scales, the noise of the
deconvolved \planck data is comparable to that of the \herschel data
(see Fig.~\ref{fig:crossCal}).  Table~\ref{tab:crossCal} shows the
cross--calibration factors for each map and wavelength, obtained as
the mean of the ratio between the \herschel and \planck visibilities
in the shaded region of Fig.~\ref{fig:crossCal}.  The
cross--calibration factors are within 20\% for every region and
wavelength.  With both datasets in the same flux scale we can now
combine them.

\subsection{Combining \planck and \herschel in the Fourier space}\label{sec:comb} 

In the last step we combine the \planck and \herschel
cross--calibrated datasets.  We generate the Fourier transforms of the
\herschel ($FT_{H}$) and \planck ($FT_{P}$) data and linearly combine
them, weighted by their correspondent \textit{uv}--scale
($\kappa = \sqrt{u^{2}+v^{2}}$) dependent functions $w_{H}(\kappa)$
and $w_{P}(\kappa)$, thus obtaining the FT of the combined image,
$FT_{C}$:
\begin{equation}\label{eq:fft-comb}
FT_{C} = FT_{H}w_{H}(\kappa) + FT^{\prime}_{P}w_{P}(\kappa),
\end{equation}
where $FT^{\prime}_{P}$ is the $FT_{P}$ after being cross--calibrated
with \herschel.

Alternatives to the classical interferometric feathering technique
that use weighting or interpolation functions different from the beam profiles have
been successfully applied when combining single dish
data~\citep[e.g., Butterworth function by][]{csengeri16}.  This
can be done because of the continuous coverage of the
\textit{uv}--plane by single dish telescopes, which allows to combine
the data at any of the overlapping scales, not limited to the
telescope beams.  In App.~\ref{sec:weights} we describe the
interpolation function used in the implementation of our method.

We define $\kappa_{eff}$ as the angular scale at which we combine the
two datasets.  To define $\kappa_{eff}$ we use Eq.~\ref{eq:fft-comb},
to define a $FT_{C}$ for each scale in the range
[$5\arcmin$,$\infty$).  For each scale, we estimate the residuals
between the new combined visibilities ($FT_{C}$) and the original
\planck and \herschel visibilities, defining $\kappa_{eff}$ as the
scale at which the residuals are minimized.  We require that
$\kappa_{eff}\geq5\arcmin$, larger than than the \planck beam.  The
zeroth Fourier mode equivalent to the constant--offset correction
occurs at $\kappa_{eff}=\infty$.  Our method is therefore a
generalization to correct the flux scales of \herschel, with the
constant--offset correction arising naturally as a special case of it
when $\kappa_{eff}=\infty$.  Table~\ref{tab:kappa-eff} we list the
$\kappa_{eff}$ values for each region and wavelength.

The very last step of our method is straightforward.  We inverse
Fourier transform the combined visibilities and the modulus of the
resulting product will be our final combined image.  

\begin{table}[t]
  \caption{Cross--calibration factors.} 
  \centering 
  \begin{tabular}{c c c c c} 
    \hline\hline 
    Region  &     & Wavelength    & [\micron]    &     \\
    			& 160 & 250           & 350          & 500\\\hline
    	HiGal--11   & 1.17$\pm$0.37 & 1.14$\pm$0.33 & 1.11$\pm$0.34 & 1.02$\pm$0.34\\
    	Per--04 & 1.19$\pm$0.32 & 1.14$\pm$0.30 & 1.08$\pm$0.31 & 1.04$\pm$0.30\\\hline
	
    \label{tab:crossCal}
  \end{tabular}\\
  \footnotesize{The errors shown in
    the cross--calibration factors correspond to the standard deviation of
    the ratio of \herschel and \planck visibilities in the scale range
    shown in Figure~\ref{fig:crossCal} and described in Section~\ref{sec:crosscal}.}
\end{table}

\begin{figure*}[t]
\includegraphics[width=0.5\textwidth]{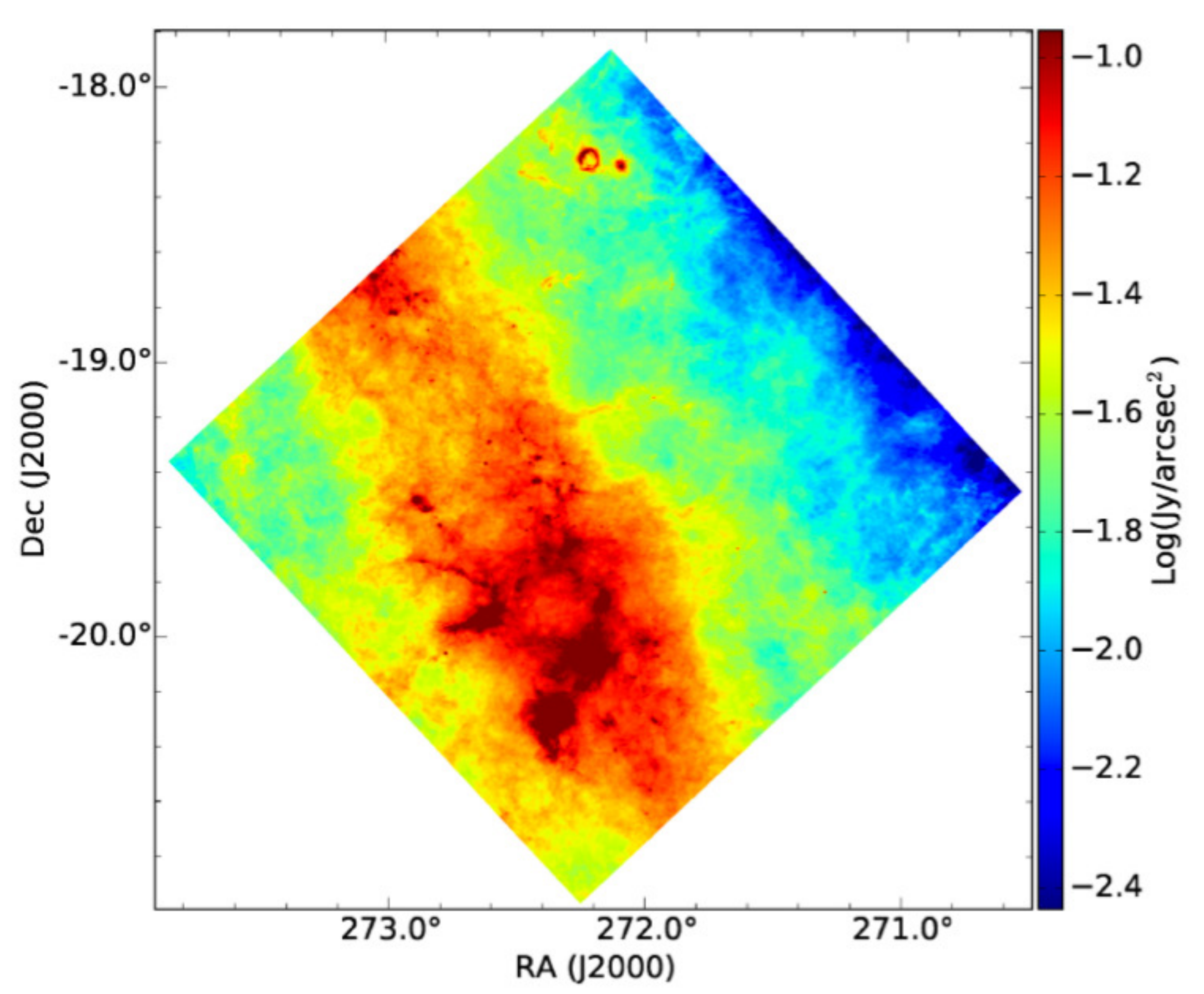}
\includegraphics[width=0.5\textwidth]{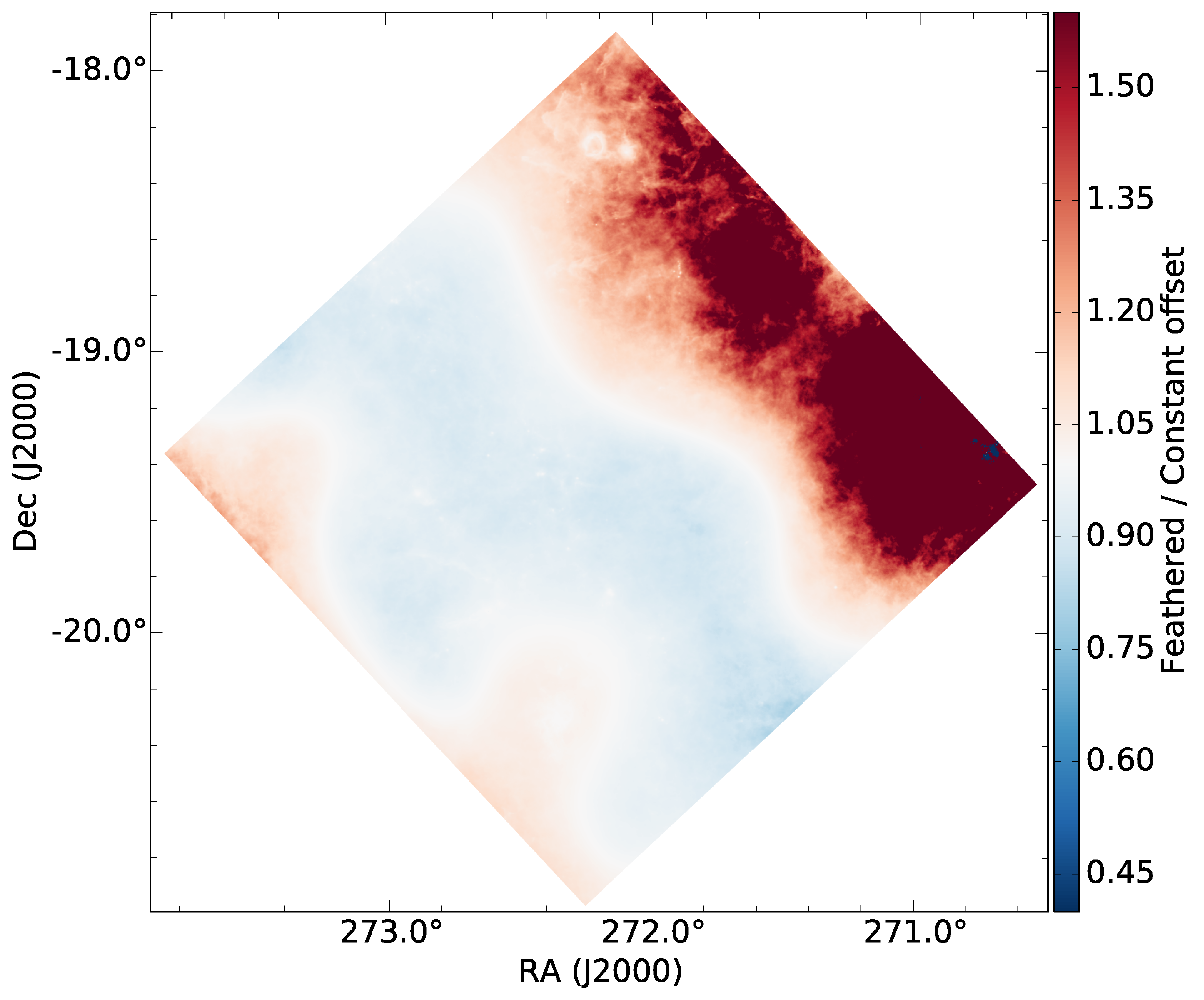}\\
\includegraphics[width=0.35\textwidth]{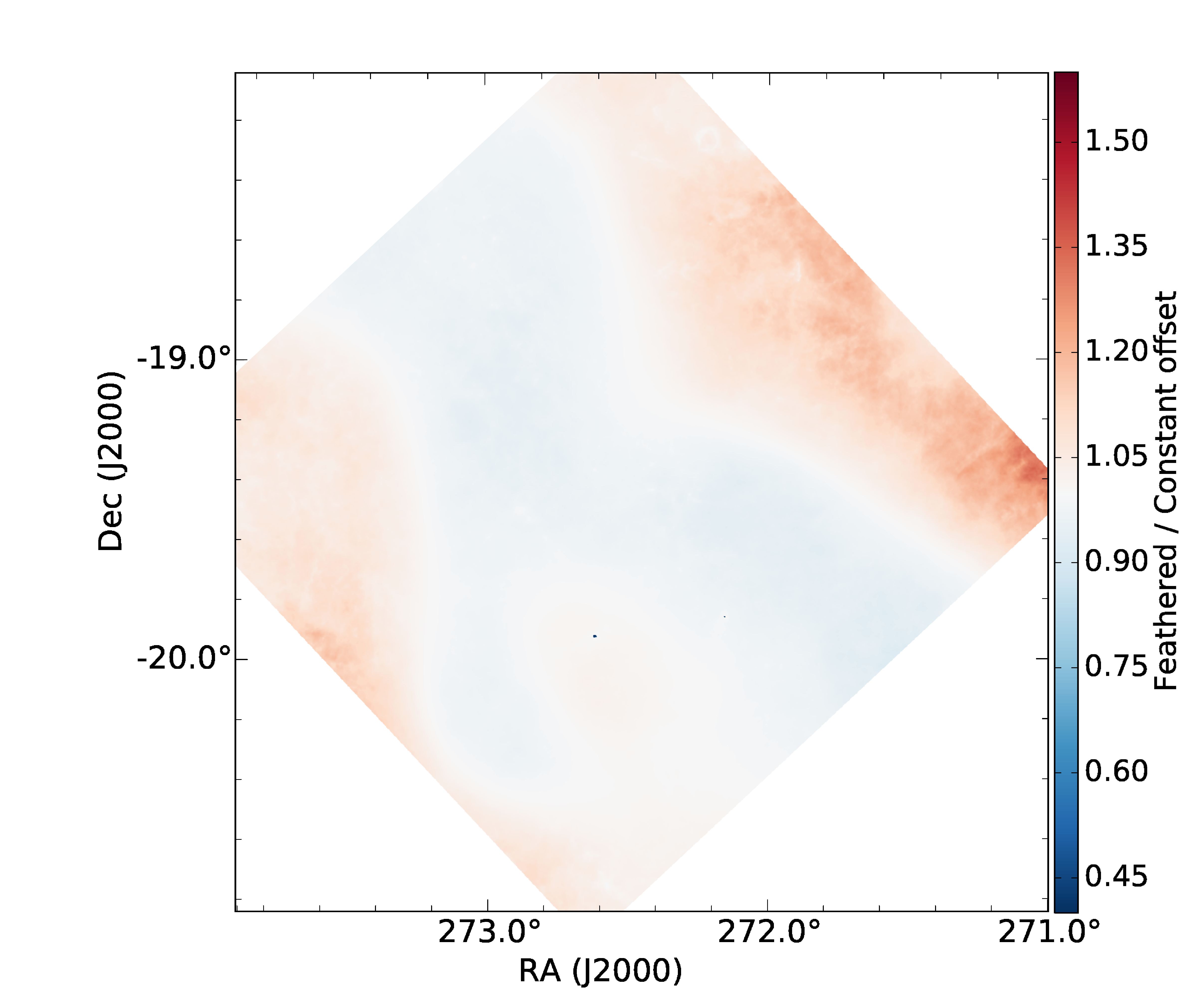}
\includegraphics[width=0.315\textwidth]{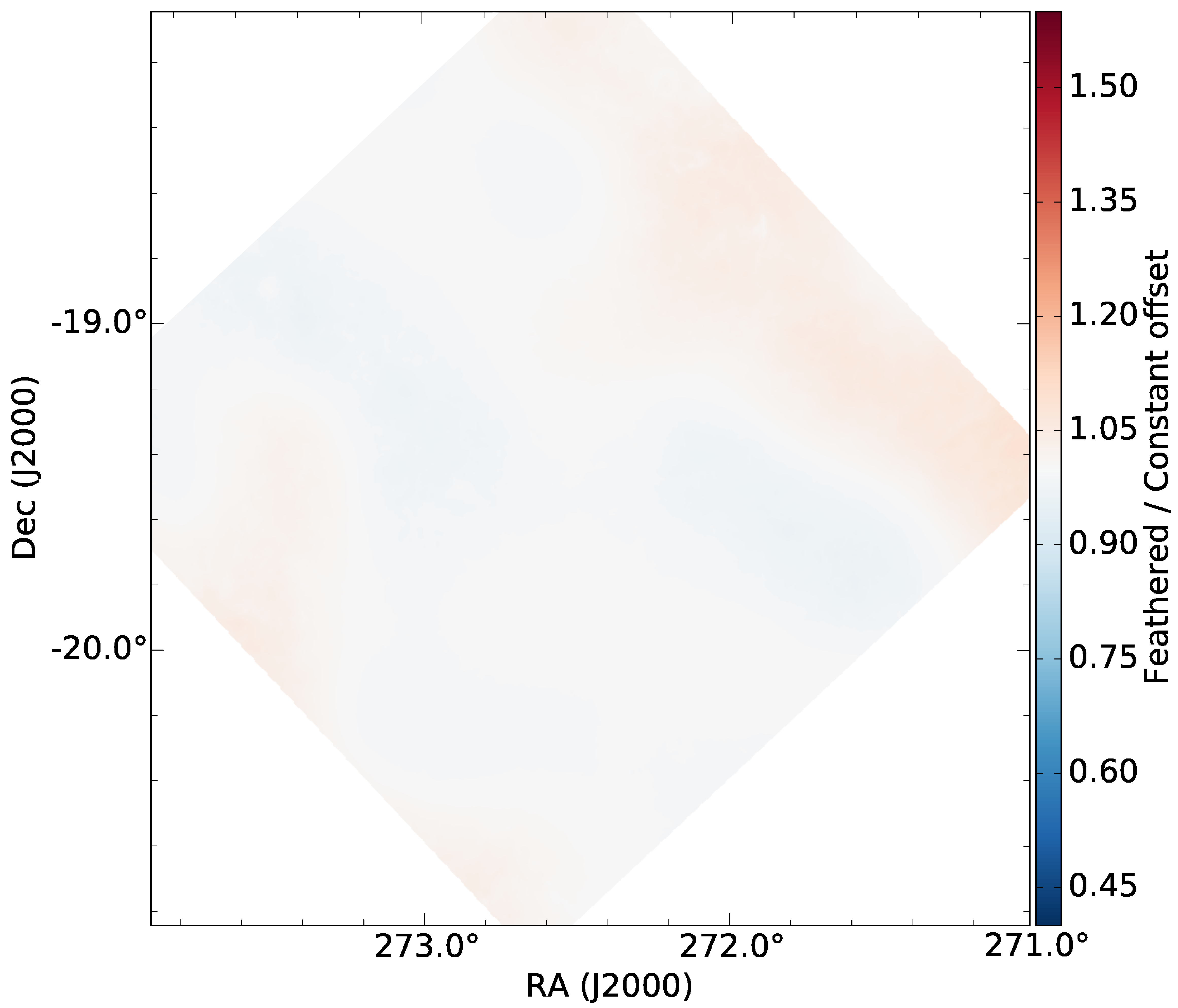}
\includegraphics[width=0.315\textwidth]{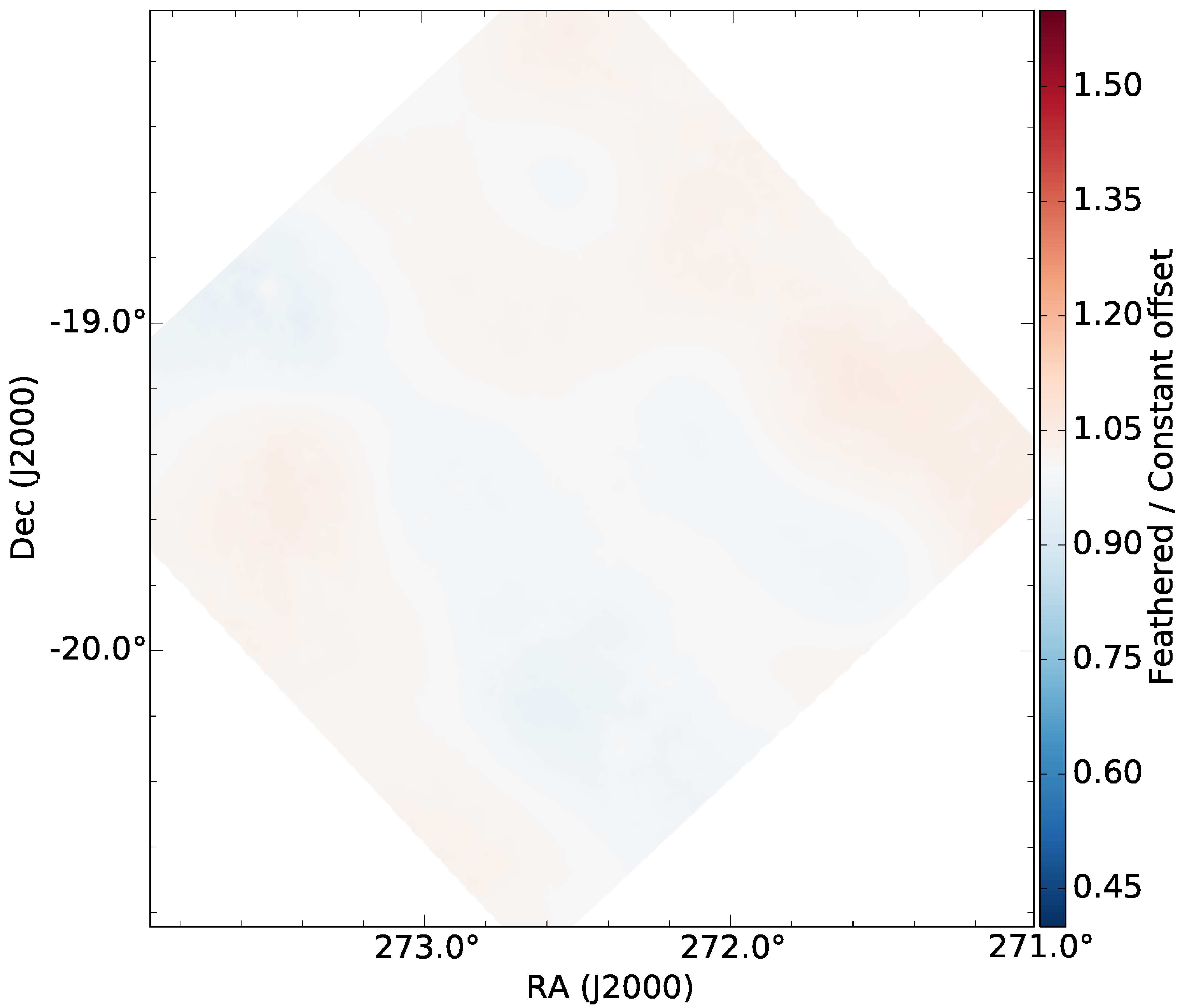}
\caption{{\it Top left:} Feathered 160~\micron image of HiGal--11, shown
    on a log scale to highlight low emission regions at large scales
    where our method has the most impact. {\it Top right:}
    Ratio of the
    feathered image over the constant--offset image.
    {\it Bottom row: } From left to right, same as in 
    the top right panel for 250\micron, 350\micron, 
    and 500\micron, showing the same colorbar in every case.}
    \label{fig:results-flux}
  \end{figure*}

\begin{figure*}[t]
\includegraphics[width=0.5\textwidth]{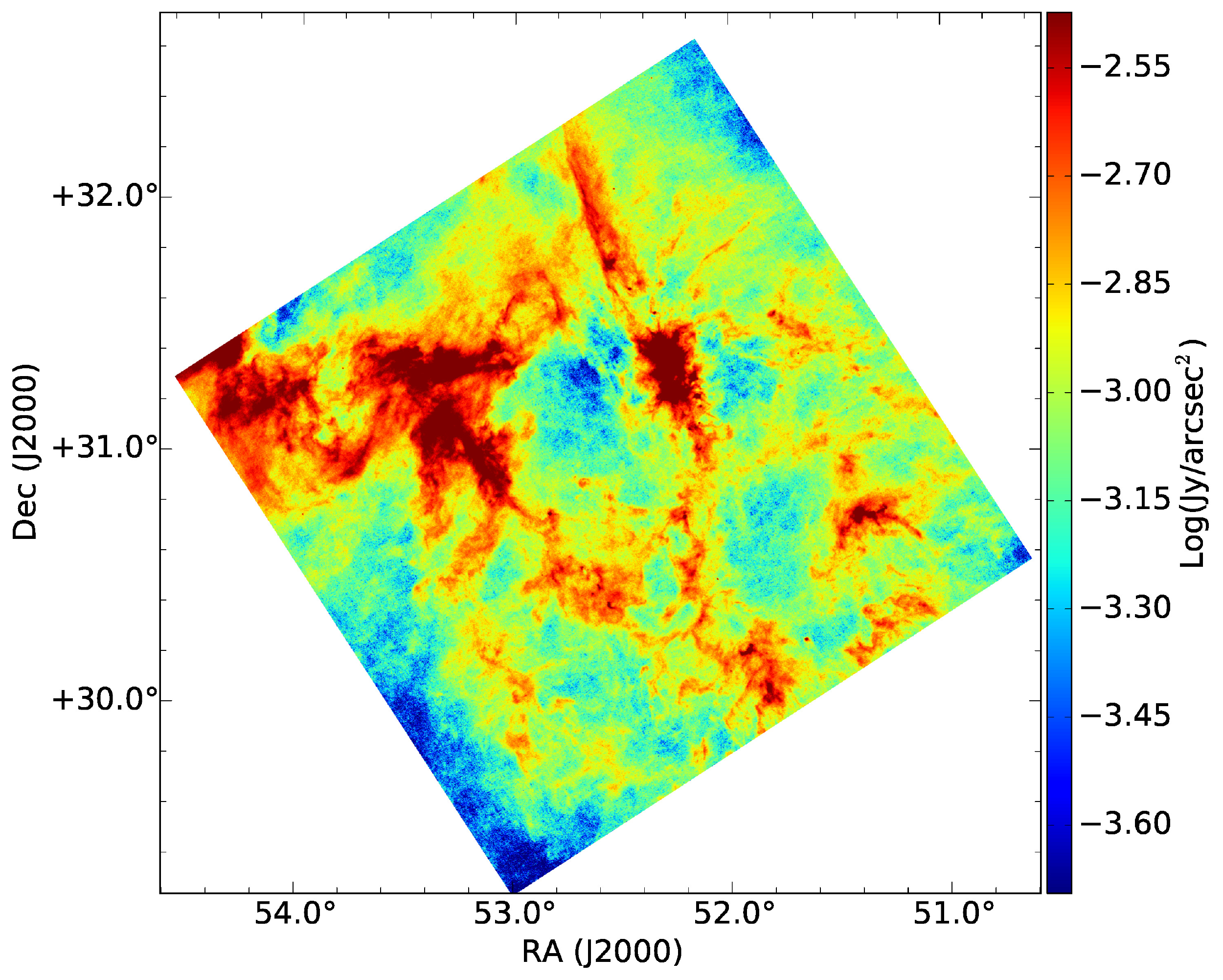}
\includegraphics[width=0.5\textwidth]{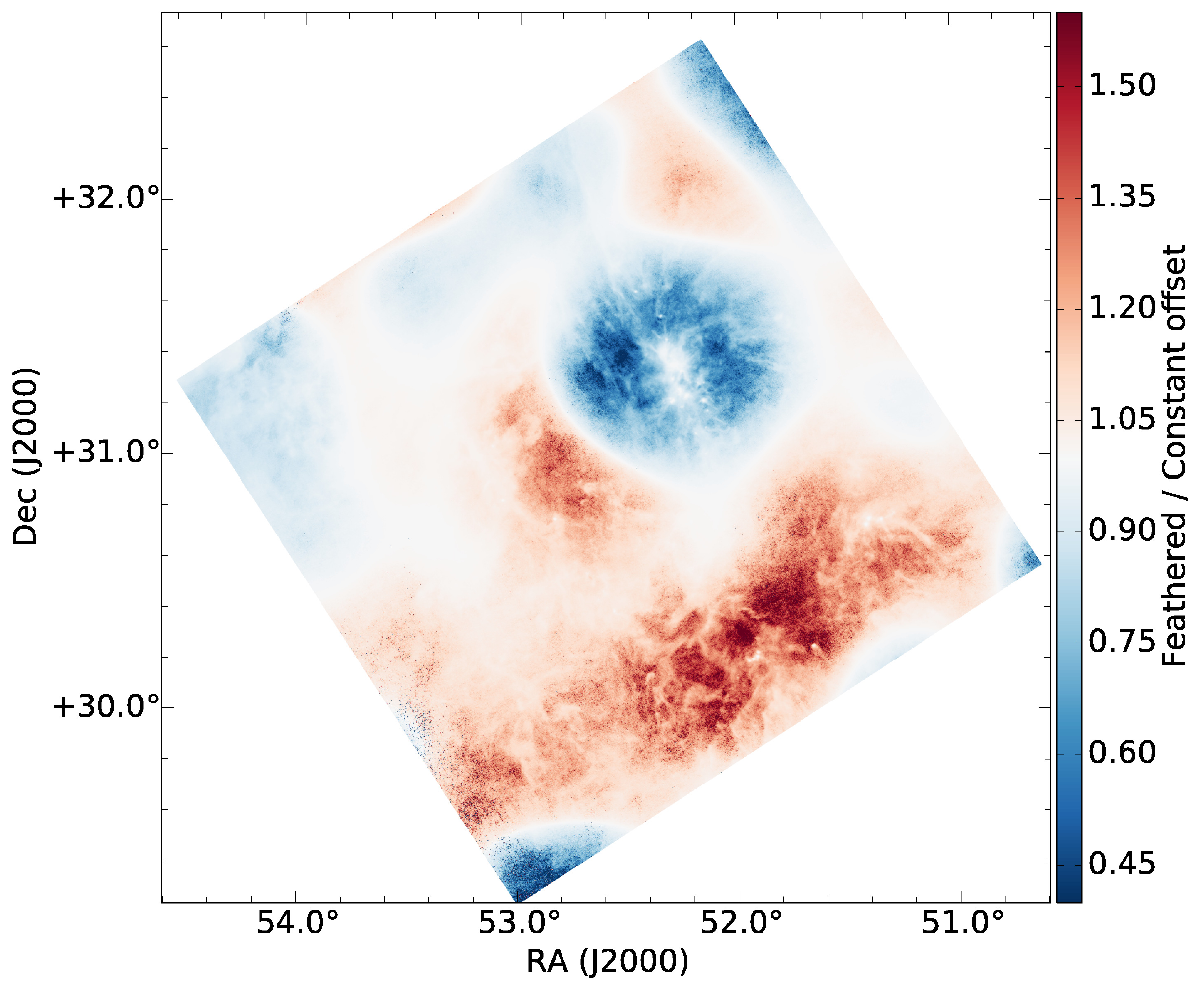}
\caption{{\it Left:} Feathered 160~\micron image of the Perseus field, shown
    on a log scale to highlight low emission regions at large scales
    where our method has the most impact. {\it Right:}
    Ratio of the
    feathered image over the constant--offset image. }
    \label{fig:results-flux-per}
  \end{figure*}

\begin{table}[t]
  \caption{Effective scale ($\kappa_{eff}$) at which \planck and 
  \herschel data are combined.} 
  \centering 
  \begin{tabular}{c c c c c} 
    \hline\hline 
    Region  &     & Wavelength    & [\micron]    &     \\
    			& 160 & 250 & 350 & 500\\\hline
    	HiGal--11   & 29.3$\arcmin$ & 23.4$\arcmin$ & 25.7$\arcmin$ & 35.1$\arcmin$\\
    	Perseus--04 & 23.4$\arcmin$ & 36.2$\arcmin$ & 15.1$\arcmin$ & 15.1$\arcmin$\\\hline
	
	\label{tab:kappa-eff}
  \end{tabular}
\end{table}

\section{Results}\label{sec:results}

Following the procedure explained in Sect.~\ref{sec:method} we
obtained the combined flux maps for each region and wavelength. In the
remainder, we will refer to these maps as ``feathered''. We show in
the top row of Fig.~\ref{fig:results-flux} the feathered map of
HiGal--11 at 160\micron on the left side and its comparison to the
constant--offset corrected map on the right side. In the bottom line
we show the comparison, for the same field, of the feathered and
constant--offset corrected maps for 250\micron, 350\micron, and
500\micron respectively. It is clear from the images that the
corrections are much more significant in for PACS (160\micron), than
for the SPIRE data. We describe this result deeper in
Sect.~\ref{sec:disc:flux}.  In Fig.~\ref{fig:results-flux-per} we show
the ``feathered'' 160\micron map of Perseus (left) and its comparison
to the constant--offset corrected map (right). For simplicity, we only
show the 160\micron case because, as in HiGal--11, it shows the most
significant differences between the ``feathered'' and constant--offset
maps. We note that the existence of strong \planck emitting sources in 
regions just outside the \herschel mapping area could in some cases
introduce artifacts near the edge of the maps. Unfortunately, such effects
would be completely random, depending exclusively on the relative
orientation between the sources and the \herschel maps. It is beyond the 
scope of this paper creating a model to quantify these effects.

We further post--processed the feathered
flux maps following the procedure in App.~\ref{sec:get-nh-t} to obtain
the feathered column density and temperature maps.  The feathered
column density map of HiGal--11 (Perseus) is shown in the top left
panel of Fig.~\ref{fig:higal-nh} (Fig.~\ref{fig:per-nh}).  The
feathered temperatures of HiGal--11 (Perseus) are shown in the top
left panel of Fig.~\ref{fig:higal-t} (Fig.~\ref{fig:per-t}).

\begin{table}[t]
  \caption{Amplitudes (in Jy/arcsec$^{2}$) of the zeroth Fourier modes of the \planck and \herschel maps.}
  \begin{center}  
    \begin{tabular}{ c | c c |  c c  }
      \hline\hline
      $\lambda$ [\micron]& HiGal--11 & & Perseus & \\
       & \herschel & \planck & \herschel & \planck \\\hline
       160 & 1.7e-2 & 3.6e-2 & 2.2e-4 &   1.3e-3 \\
       250 & 2.3e-2 & 2.4e-2 & 1.2e-3 &   1.2e-3 \\
       350 & 1.1e-2 & 1.1e-2 & 7.5e-3 &   7.5e-3 \\
       500 & 3.9e-3 & 3.9e-3 & 3.2e-4 &   3.4e-4 \\\hline
    \end{tabular}
  \end{center}
  \label{tab:constant-ofset}
\end{table}

\section{Discussion}\label{sec:disc}

In this section we first compare our feathered flux maps with those
obtained using the constant--offset correction. We then compare the
feathered and constant--offset column densities and temperatures. 

\subsection{Feathered vs constant--offset flux maps}\label{sec:disc:flux}

We now compare the our feathered flux maps with those 
obtained via the constant--offset correction applied 
in previous works~\citep[e.g.,][]{bernard10,lombardi14,zari16}.
In Fig.~\ref{fig:results-flux} we show the comparison 
between our feathered HiGal--11 flux maps and
the constant--offset maps. 
The feathered image of HiGal--11 has more emission in the diffuse
regions off of the Galactic plane, and the relative differences can 
exceed 40\% (although the absolute differences are similar)
over significant areas of the image. Contrary, the feathered image
tends to show $~\sim10\%$ lower fluxes on the Galactic plane areas.
The feathered and constant--offset images agree on compact objects.
The map of Perseus exhibits a similar behavior to that of
the HiGal--11 field, with the most strong emitting regions
having similar fluxes in the feathered and constant--offset maps
and the constant offset map over--estimating (under--)
the flux at intermediate (low) fluxes (see Fig.~\ref{fig:results-flux-per}).
These results illustrate non--uniform and scale 
dependent nature of the signal at large scales, specially significant
in the PACS data.

In the bottom row of Fig.~\ref{fig:results-flux} we show the 
comparison between the feathered and constant--offset flux
maps for the same region for the SPIRE wavelengths. In general, 
the feathered and constant--offset agree within calibration
errors at these wavelengths, as it can be seen in the predominantly 
white maps in the bottom row of Fig.~\ref{fig:results-flux}. Only
small parts of the 250\micron image show significant differences
between the feathered and constant--offset maps. This result
agrees with~\citet{bertincourt16} who study the large scale emission
of SPIRE and \planck in several regions finding good agreement
between both. For this reason, and to simplify the paper, we do not
show the comparison between the feathered and constant--offset
corrected maps for Perseus at the SPIRE wavelengths, 
since the results are similar to those already described.

\begin{figure*}
\includegraphics[width=0.48\textwidth]{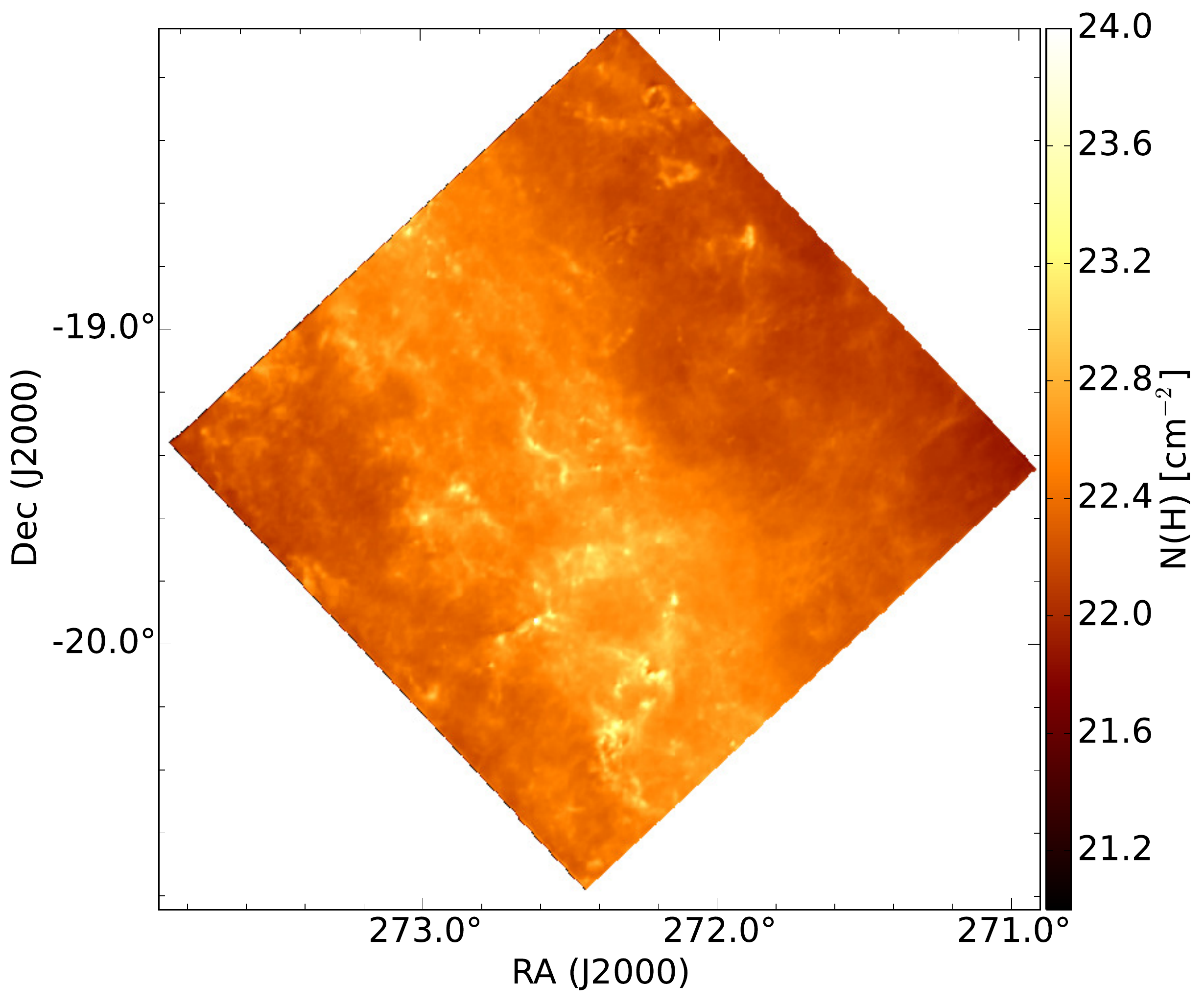}
\includegraphics[width=0.48\textwidth]{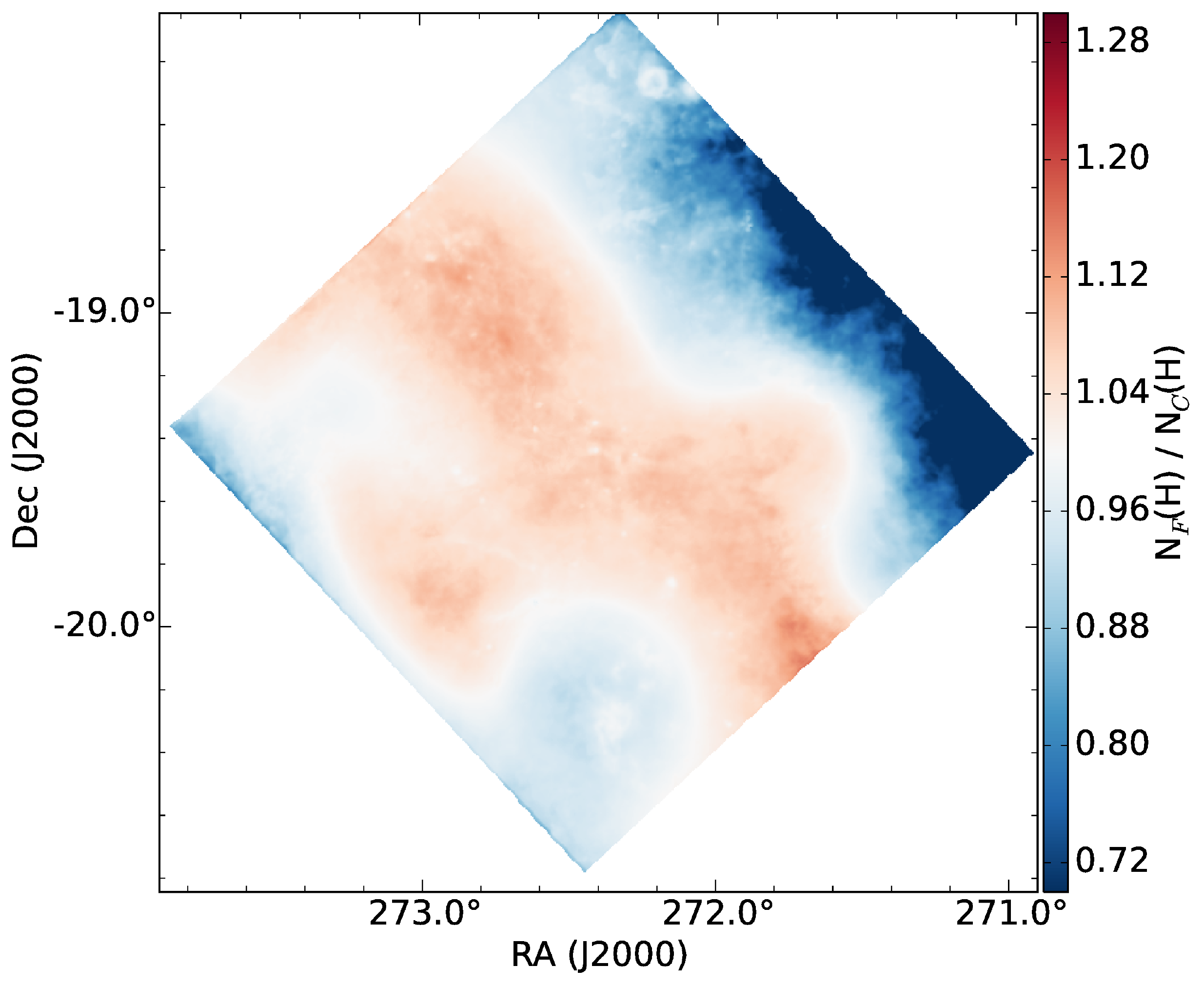}\\
\includegraphics[width=0.45\textwidth]{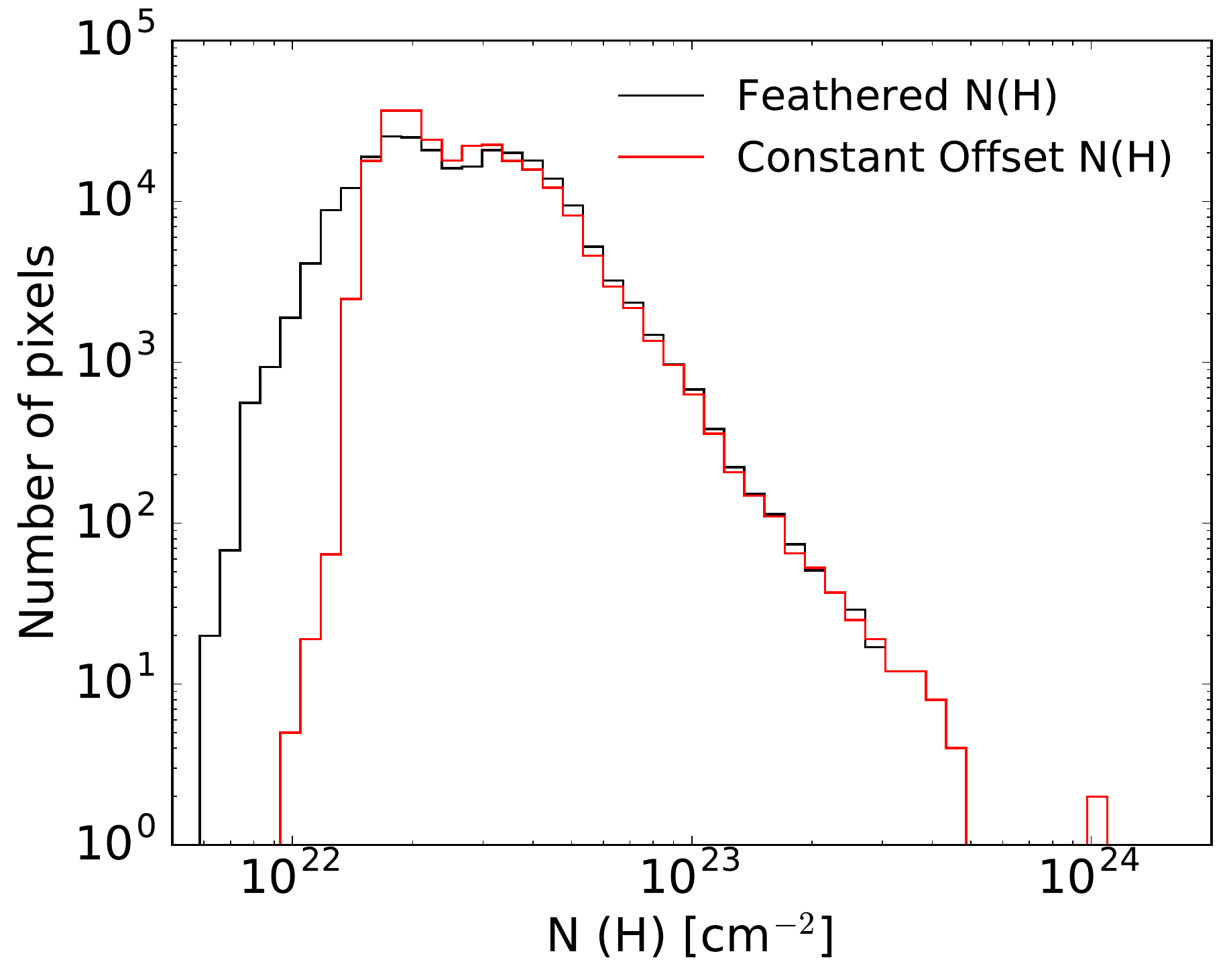}\,\,\,\,\,\,\,\,\,\,\,\,\,
\includegraphics[width=0.55\textwidth]{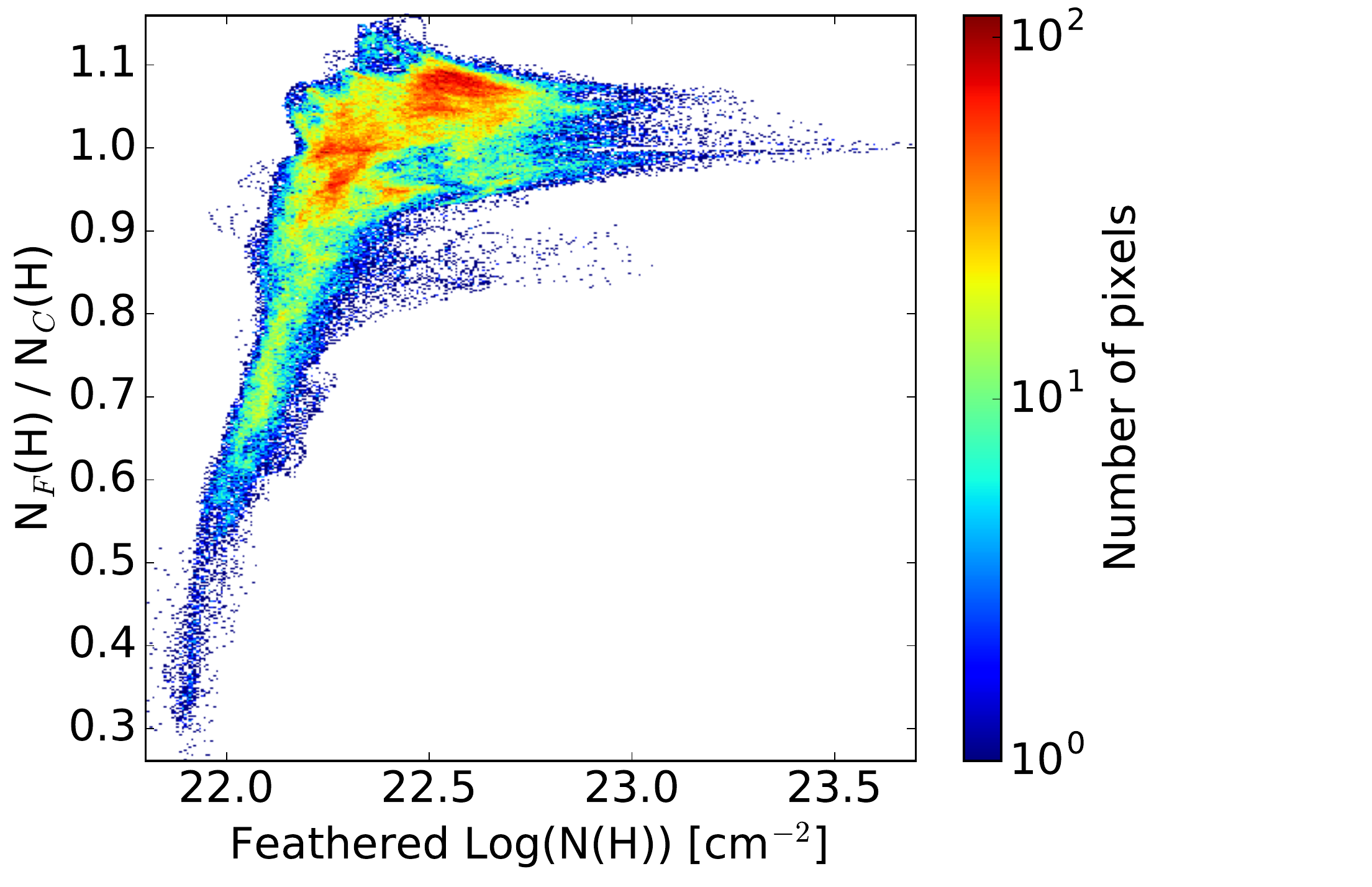}
\caption{\textit{Top left:}  Logarithmic column density map of the HiGal--11 obtained with our method.
\textit{Top right: } Ratio of our feathered and the constant--offset
column density maps.
\textit{Bottom left: Histograms of the feathered (black) and the 
constant--offset (red) column density maps.}
\textit{Bottom right: } Ratio of feathered and
constant--offset corrected column densities $N_{F}(H)/N_{C}(H)$
function of the feathered column density.}
  \label{fig:higal-nh}
\end{figure*}

\begin{figure*}
\includegraphics[width=0.48\textwidth]{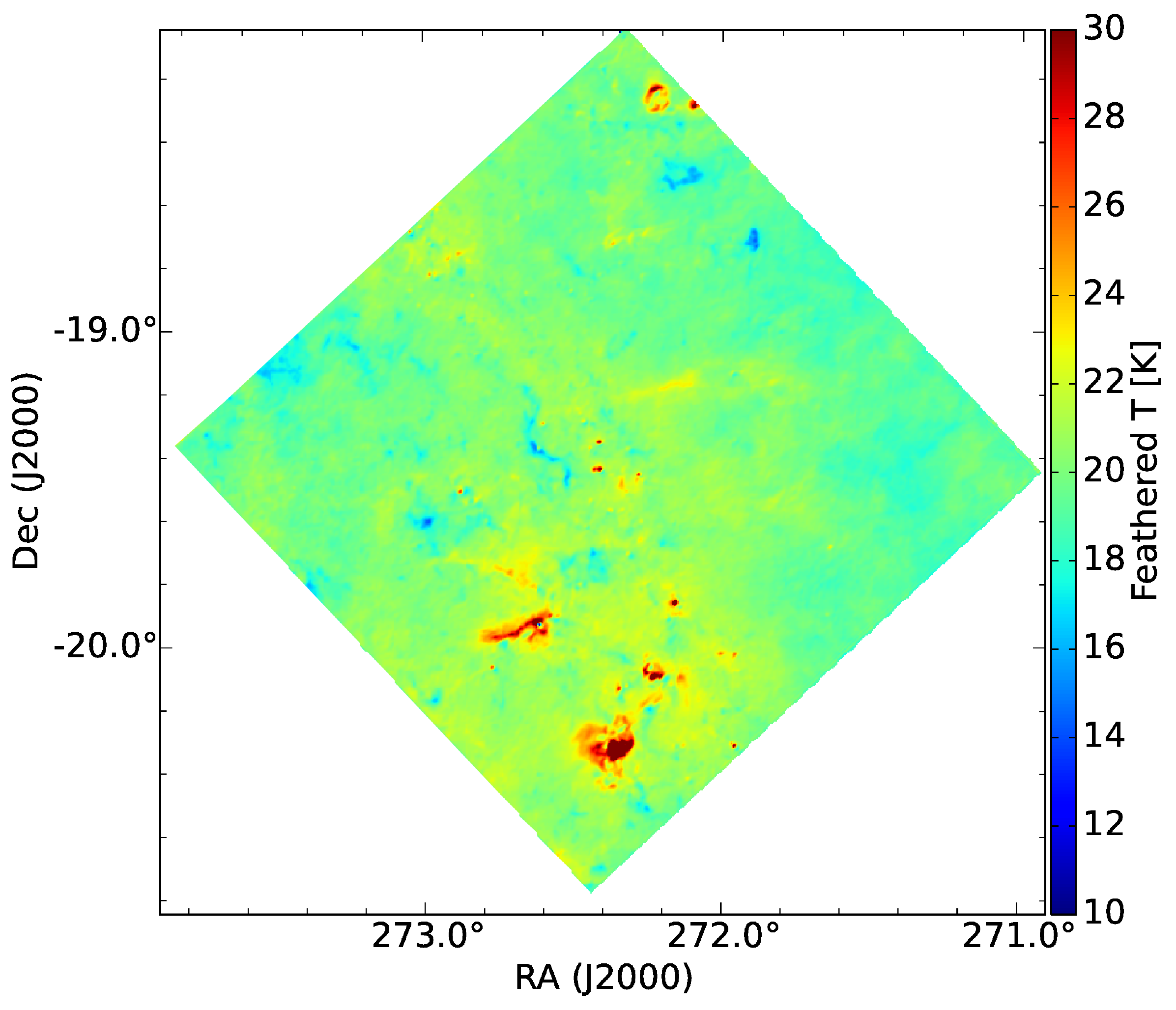}
\includegraphics[width=0.48\textwidth]{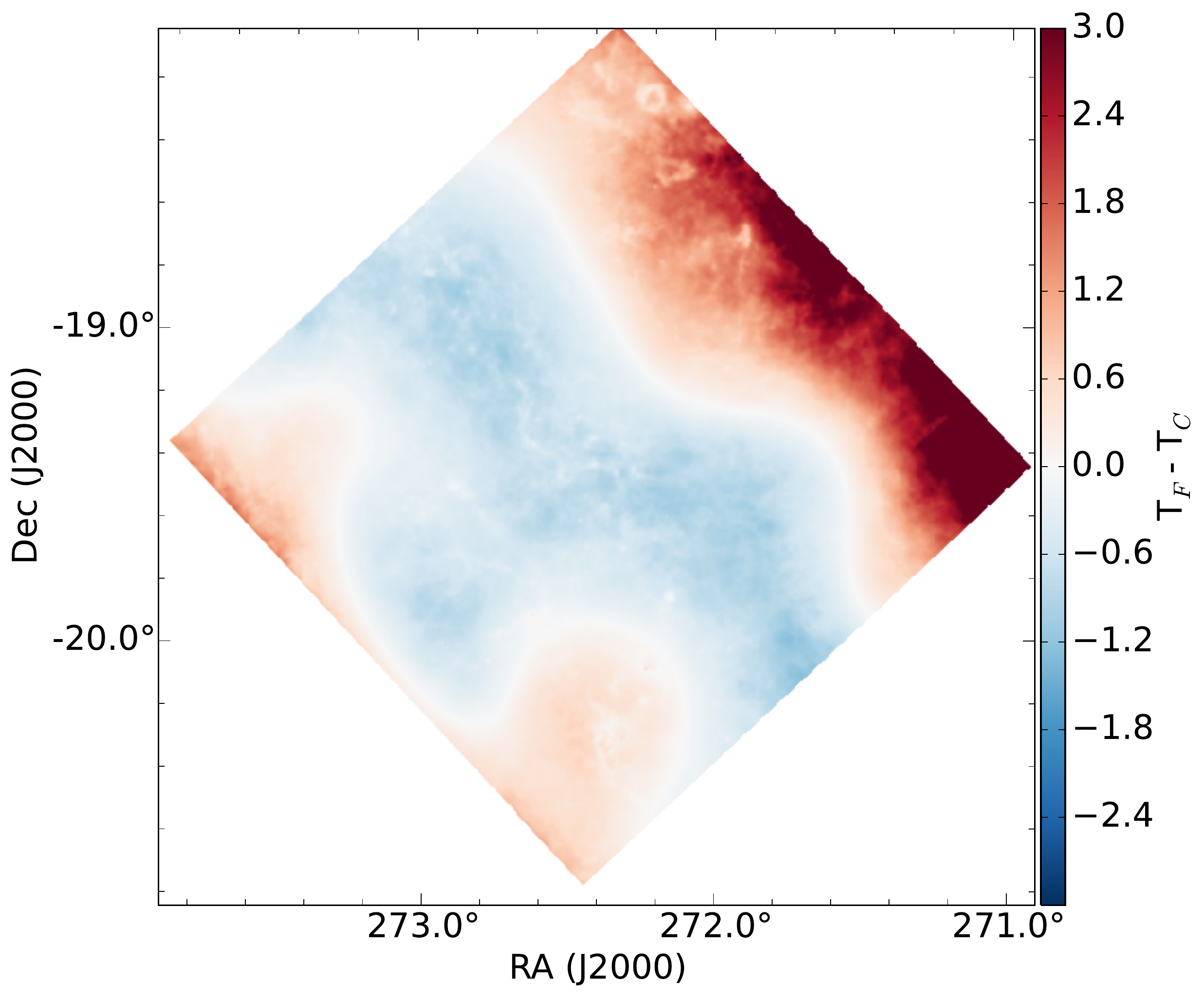}
\includegraphics[width=0.45\textwidth]{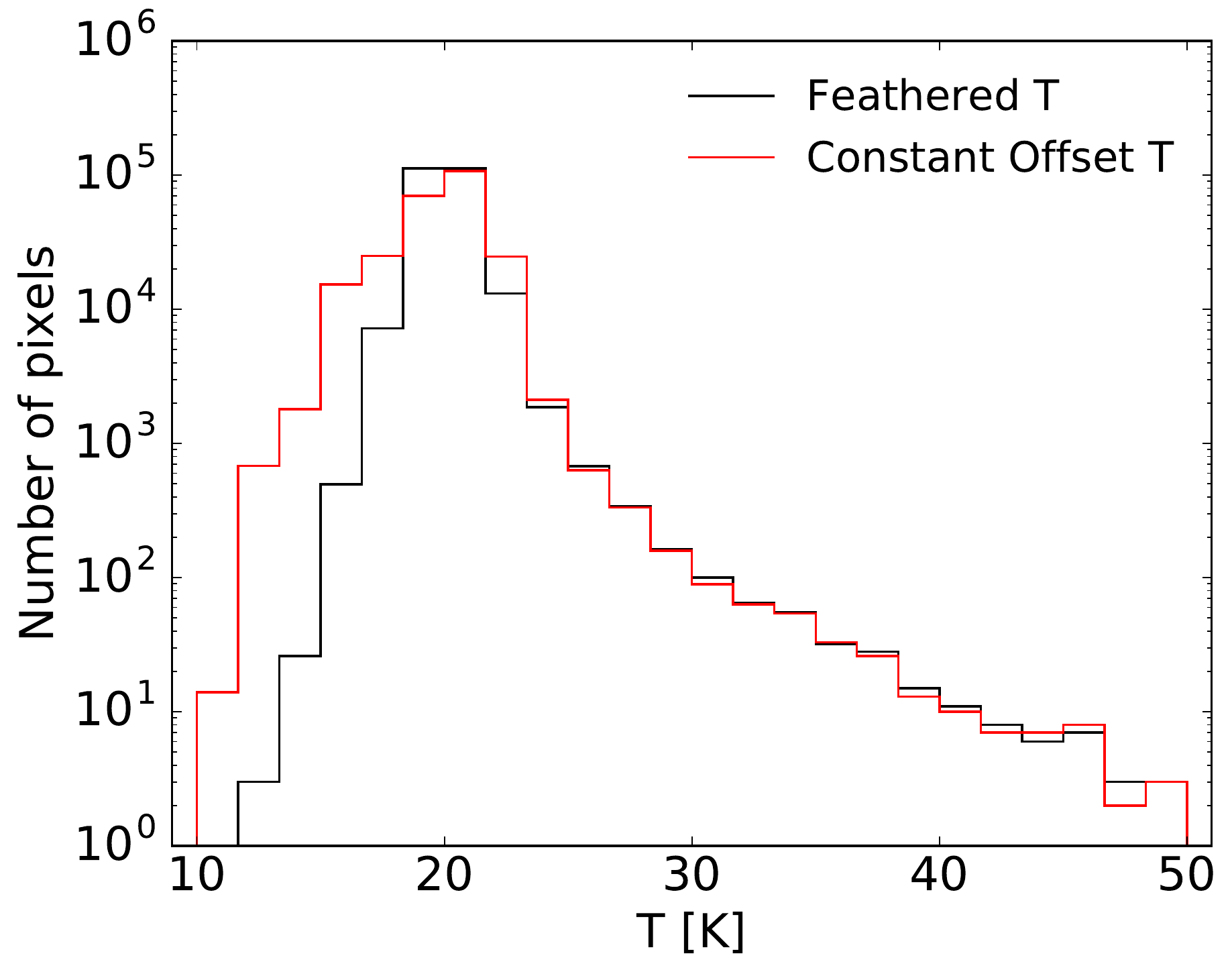}\,\,\,\,\,\,\,\,\,\,\,\,\,\,\,\,\,\,\,\,\,
\includegraphics[width=0.9\textwidth]{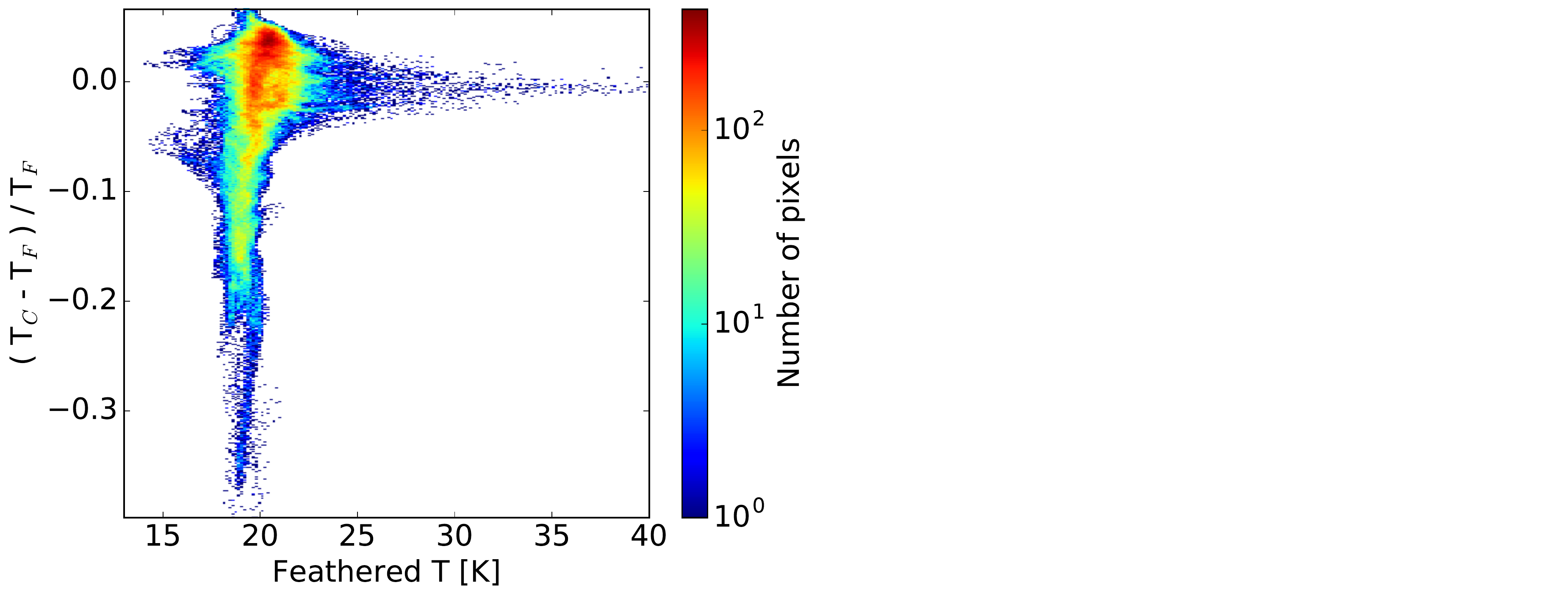}
\caption{\textit{Top left:} Temperature map of the HiGal field
11 obtained with our method.
\textit{Top right: } Difference of our feathered and the constant--offset
temperature maps.
\textit{Bottom left: }Histograms of the feathered (black) and the 
constant--offset (red) temperature maps.
\textit{Bottom right: } Residuals, ($T_{C}-T_{F})/T_{F}$, of the feathered and constant--offset
temperature maps as function of the feathered temperature.}
  \label{fig:higal-t}
\end{figure*}

\begin{figure*}
\includegraphics[width=0.48\textwidth]{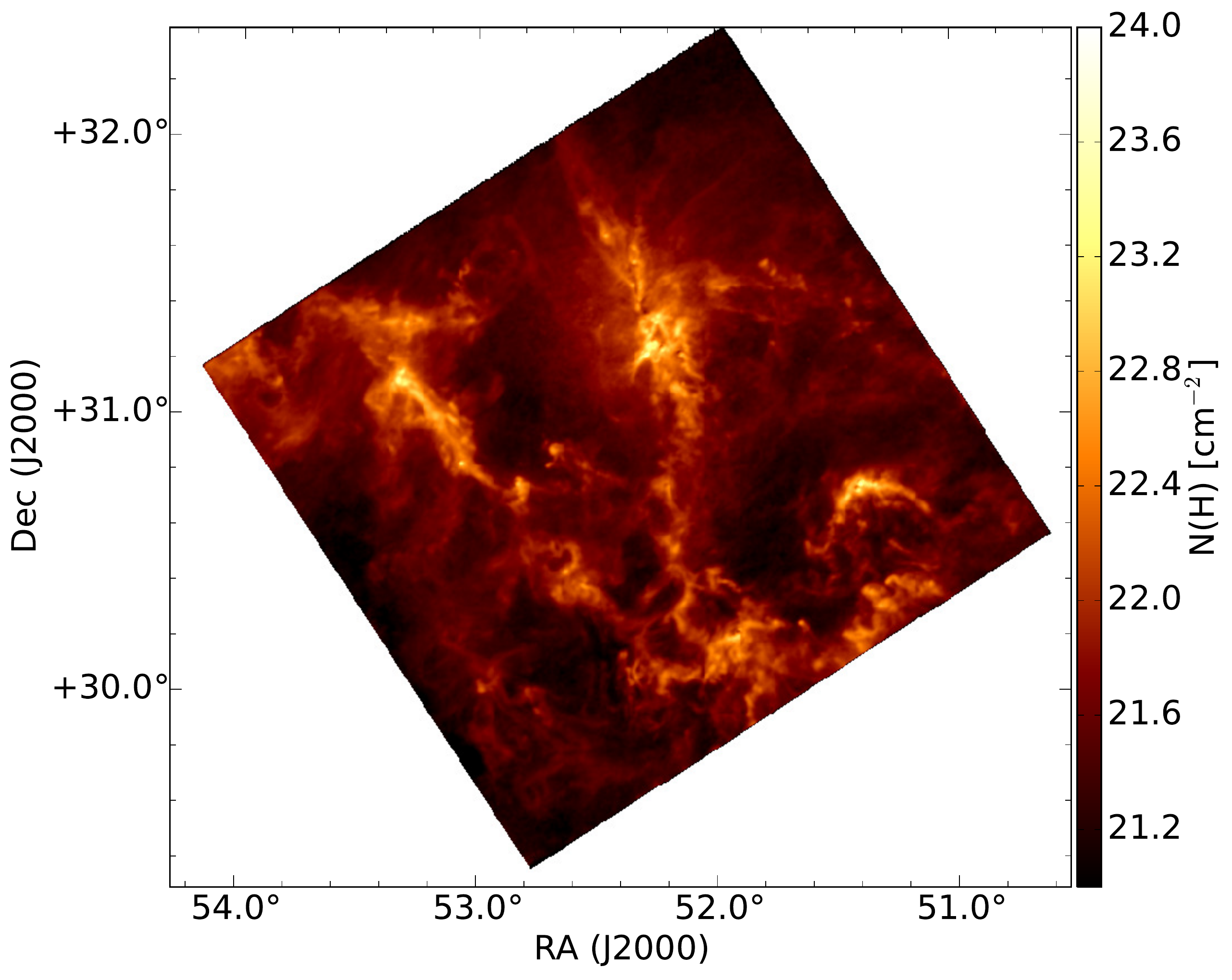}
\includegraphics[width=0.48\textwidth]{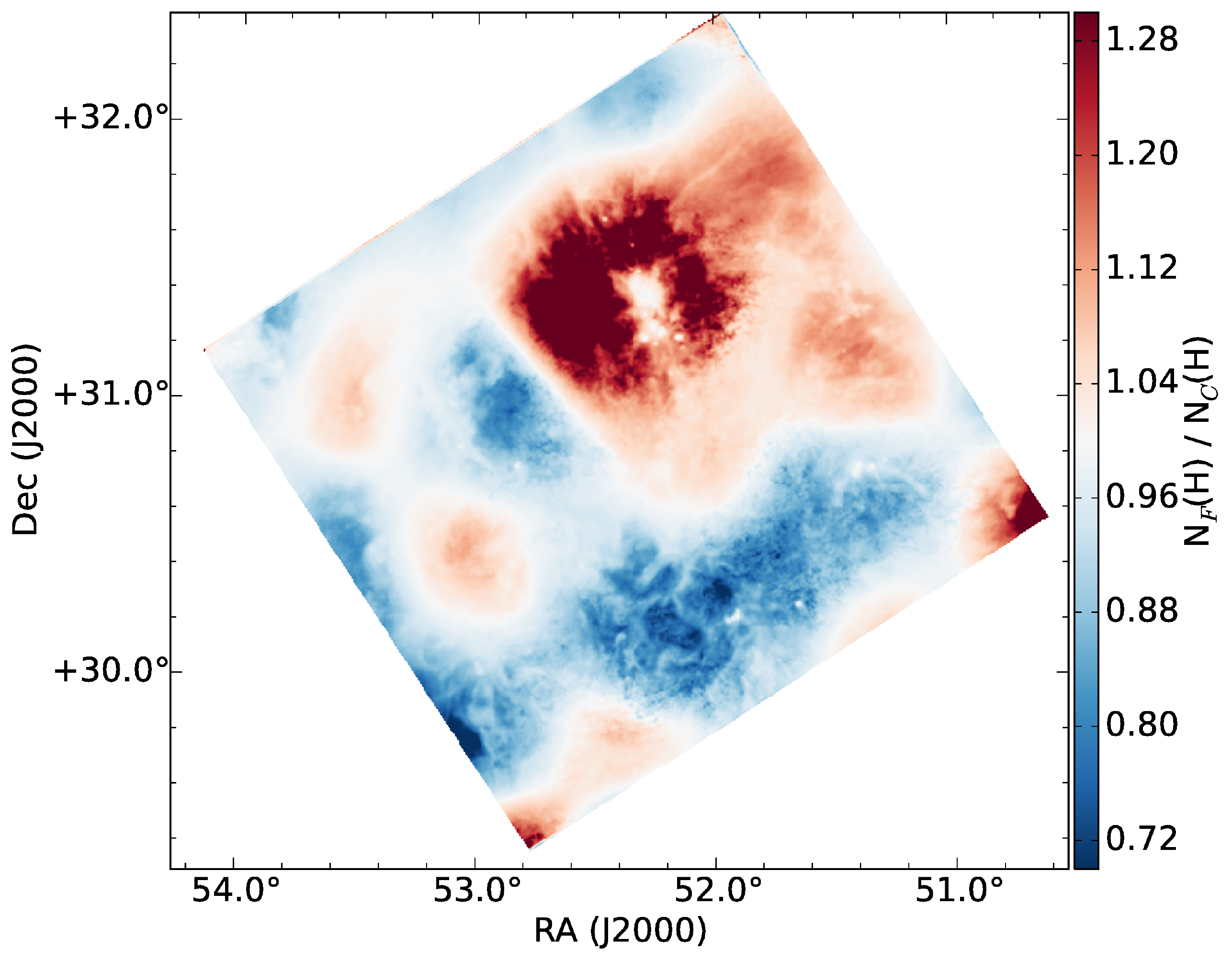}\\
\includegraphics[width=0.45\textwidth]{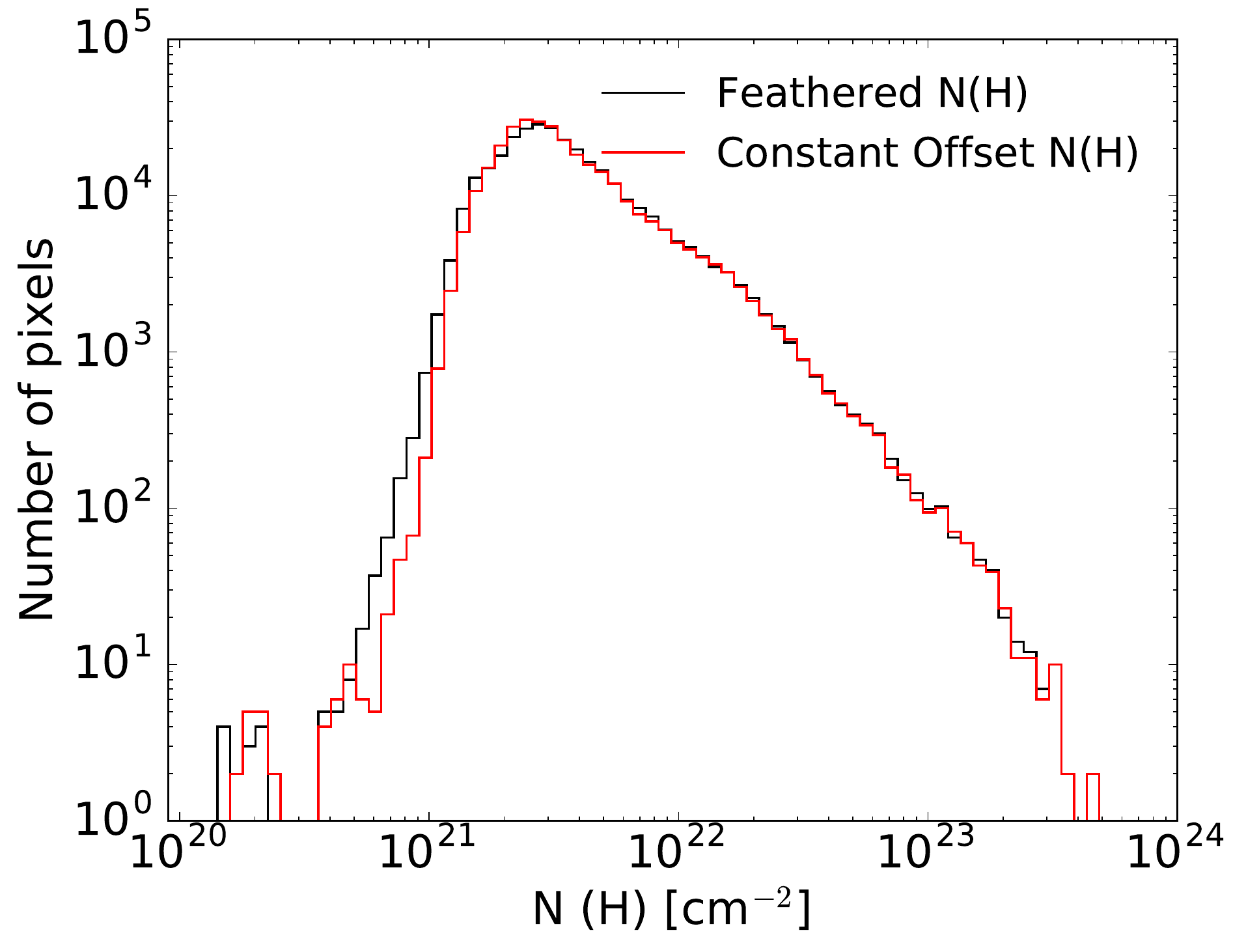}\,\,\,\,\,\,\,\,\,\,\,\,\,
\includegraphics[width=0.45\textwidth]{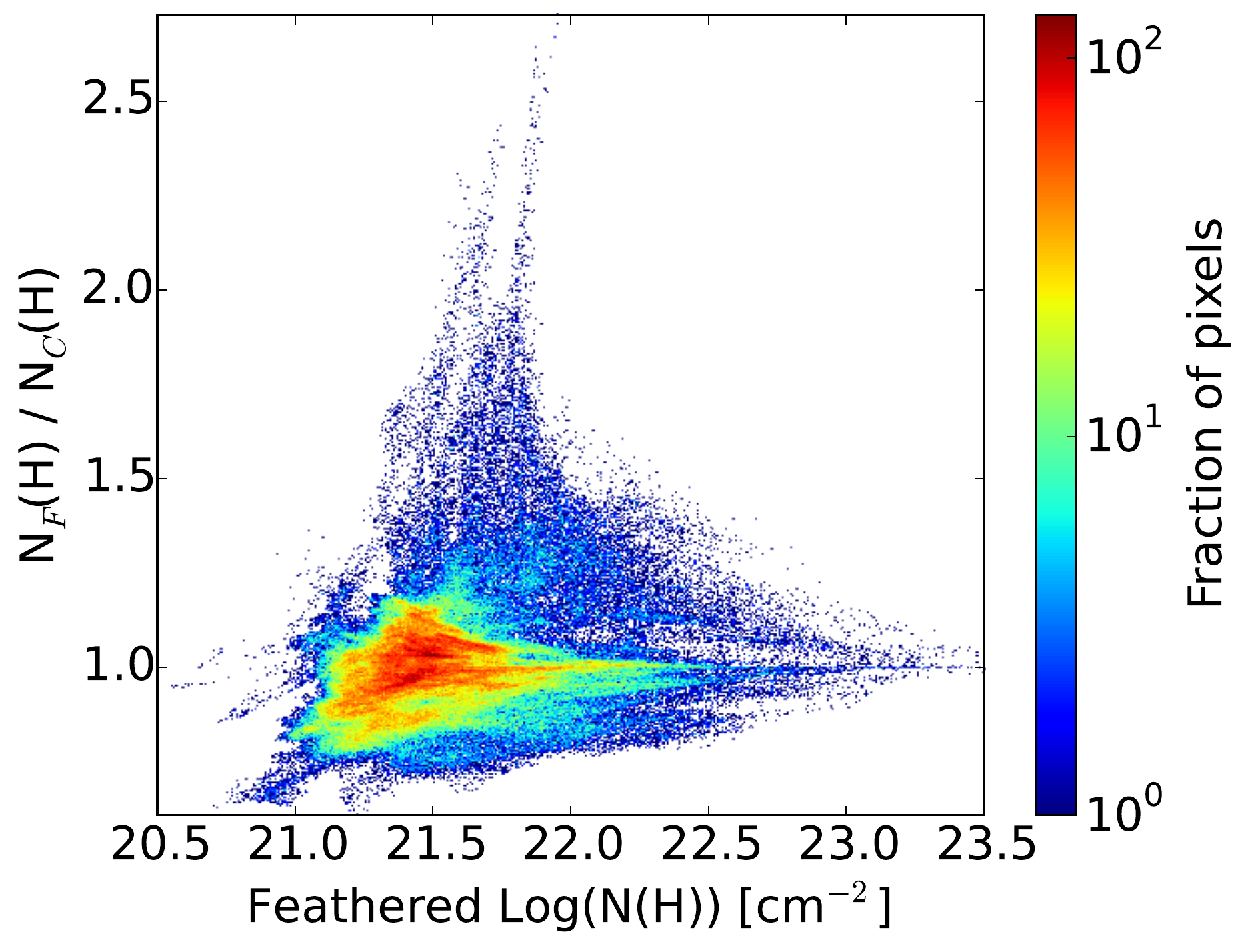}
\caption{Same as Fig.~\ref{fig:higal-nh} for Perseus.}
  \label{fig:per-nh}
\end{figure*}

\begin{figure*}
\includegraphics[width=0.48\textwidth]{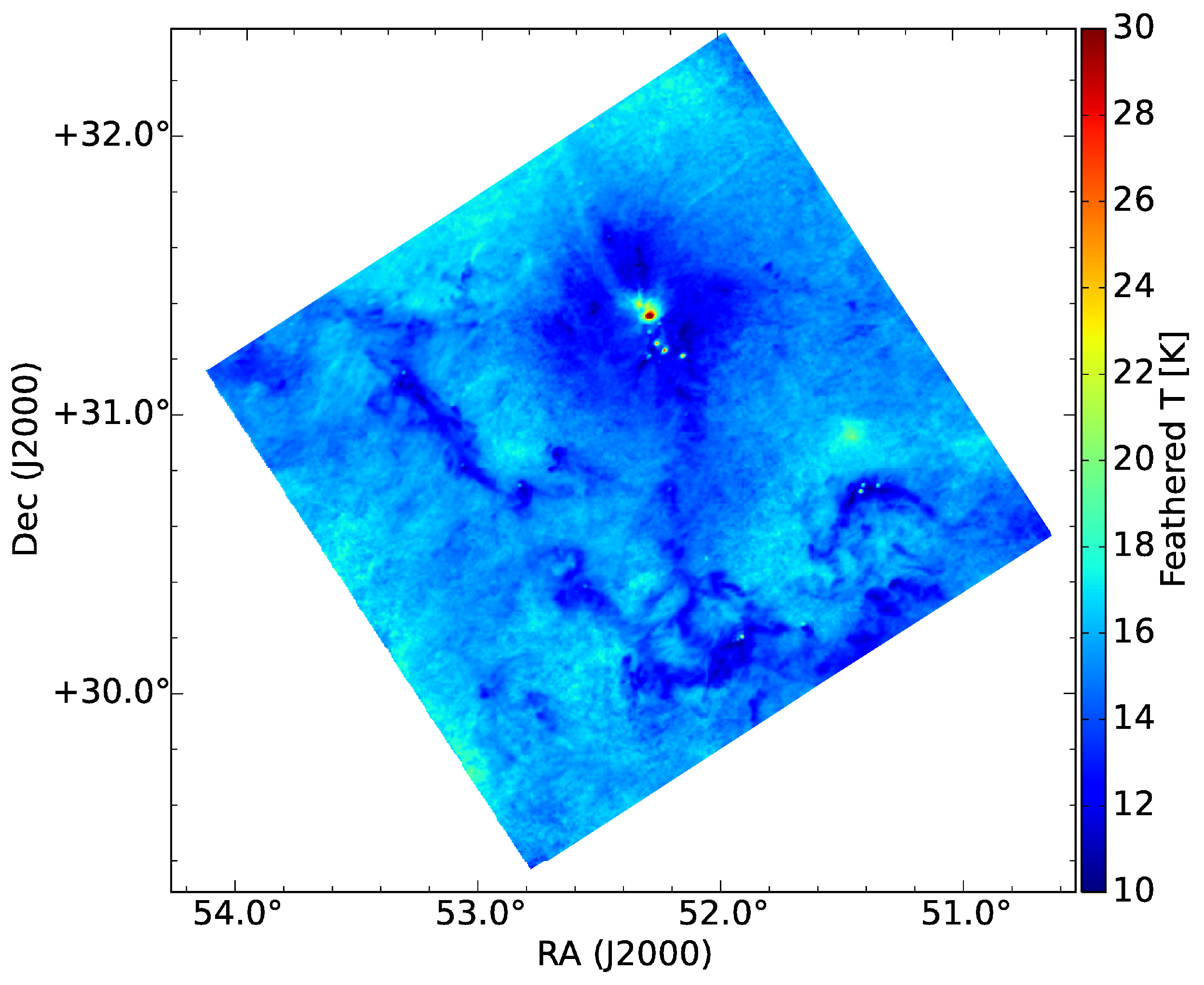}
\includegraphics[width=0.48\textwidth]{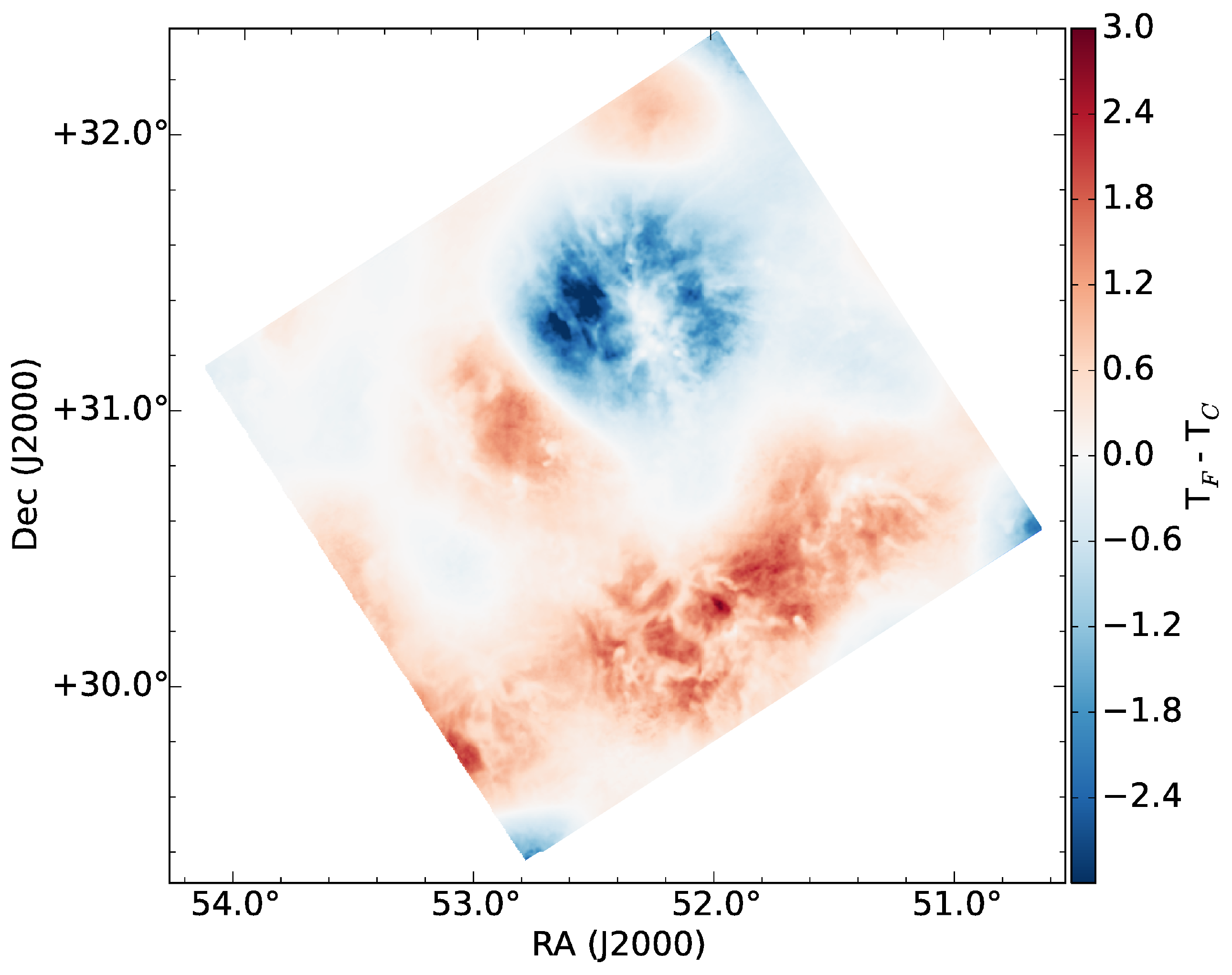}
\includegraphics[width=0.45\textwidth]{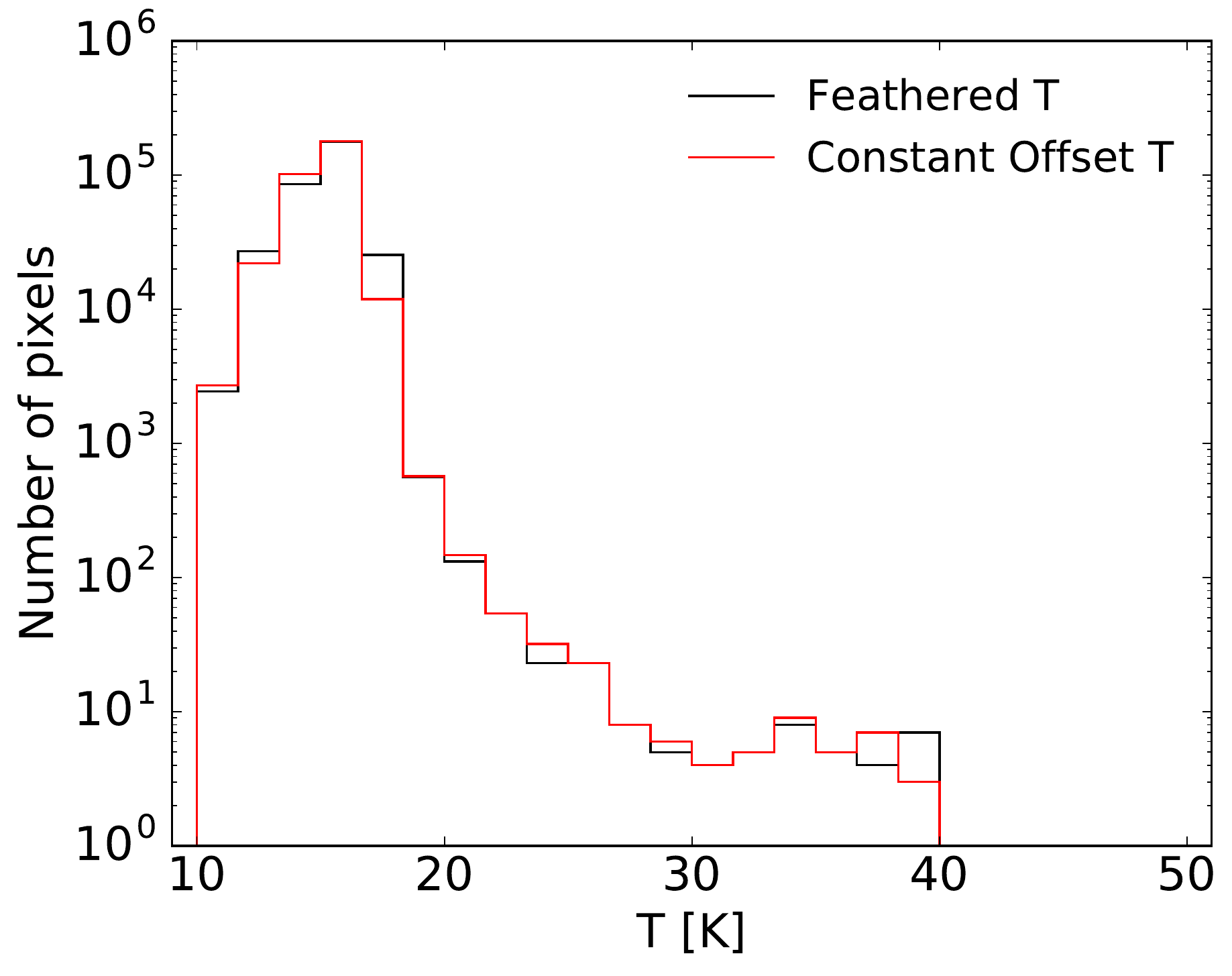}\,\,\,\,\,\,\,\,\,\,\,\,\,\,\,\,\,\,\,\,\,
\includegraphics[width=0.78\textwidth]{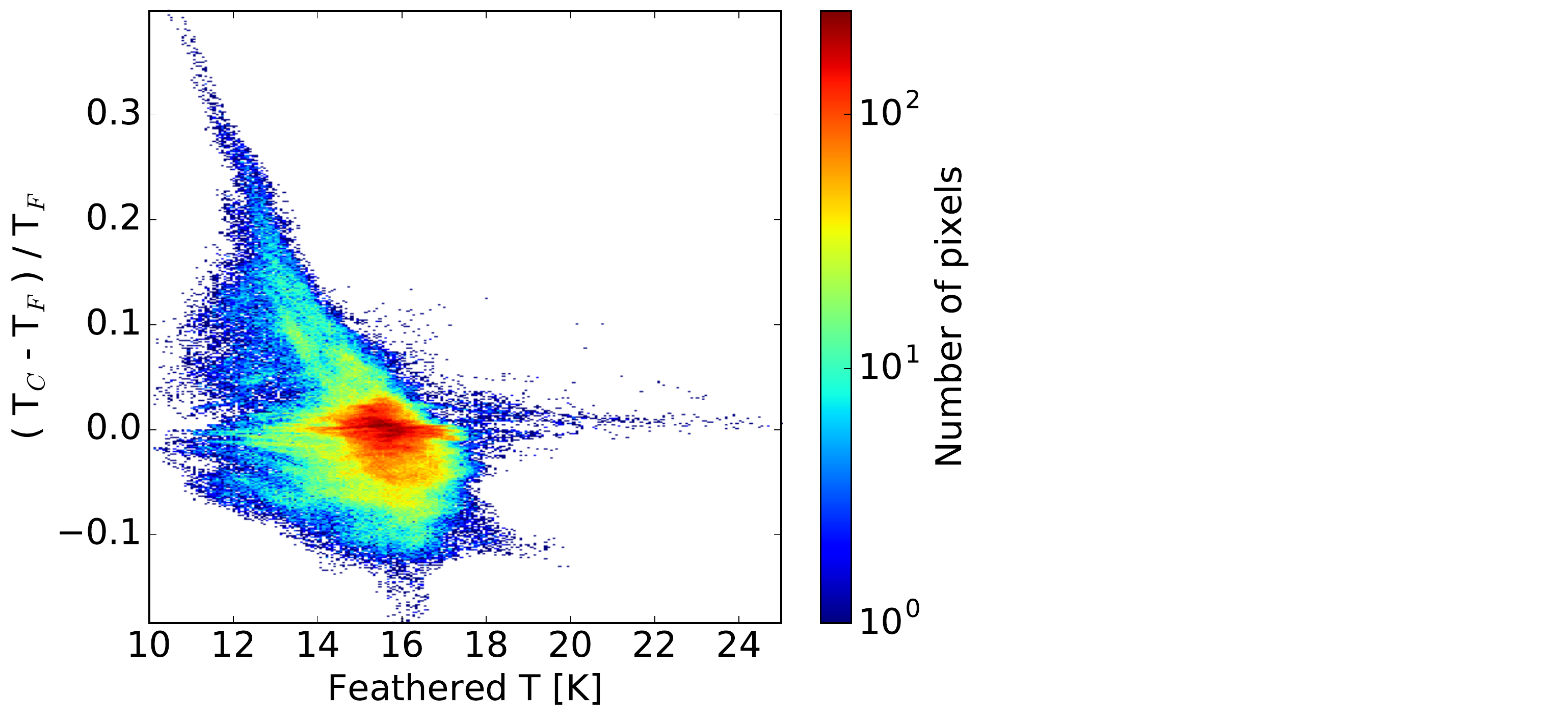}
\caption{Same as Fig.~\ref{fig:higal-t} for Perseus.}
  \label{fig:per-t}
\end{figure*}

\subsection{Comparing column density and temperature maps to previous methods}\label{sec:compare}

We have shown in Sect.~\ref{sec:disc:flux} 
that the \herschel data corrected with a constant--offset tend
to underestimate the fluxes in diffuse regions compared
to our flux feathered maps, while both agree well in regions 
with strong emission. These results are 
specially significant at 160\,\micron. 
With these results we would expect that 
the temperatures in diffuse regions are underestimated
by \herschel and therefore the column densities 
overestimated.

This is exactly what we find in HiGal--11 and Perseus, as it is
shown in Fig.~\ref{fig:higal-nh}, Fig.~\ref{fig:higal-t},
Fig.~\ref{fig:per-nh}, and Fig.~\ref{fig:per-t}. In strong 
emitting (i.e. dense) regions, the feathered and
constant--offset maps agree for both, temperatures
and column densities. This is shown shown with the white regions
in the ratio map, the similar high column density tails of the
histograms, and the surface density points follow the identity
at large column densities (and low temperatures). 
In HiGal--11 constant--offset maps do not
measure column densities lower than 10$^{22}$\,cm$^{-2}$. 
The inverse effect is seen in temperatures:
the constant--offset temperatures below 20\,K tend to be 
significantly lower than our feathered temperatures 
(see Fig.~\ref{fig:higal-t}). 
The map shows discrepancies larger than 30\% 
between both methods over $\sim15\%$ of the area of HiGal--11.
In the case of Perseus the temperature difference map shows
that in general, the constant--offset and feathered temperatures
agree within 2\,K in Perseus. The differences in column densities
are concentrated on the gas surrounding the NGC 1333 region.
These differences account for more than 30\% at intermediate
(10$^{22}$\,cm$^{-2}$) column densities. 

These results highlight the importance of a proper treatment of the 
\herschel data, specially in diffuse regions, since the column densities
are directly related with the mass of the dust, and therefore
the total mass of the molecular clouds, intimately linked
to key physical parameters as the gravitational potential. 

\section{Testing our method on simulations}\label{sec:sims}

\begin{figure*}[t]
\includegraphics[width=0.5\textwidth]{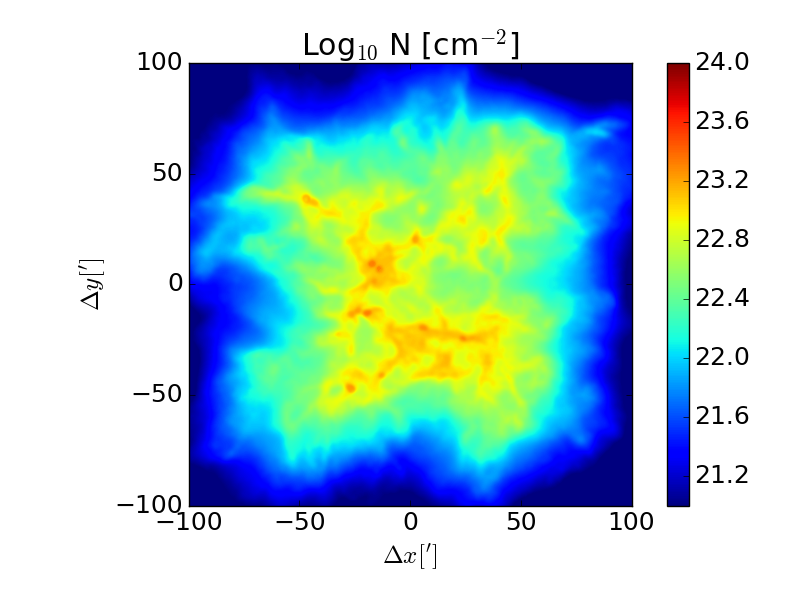}
\includegraphics[width=0.5\textwidth]{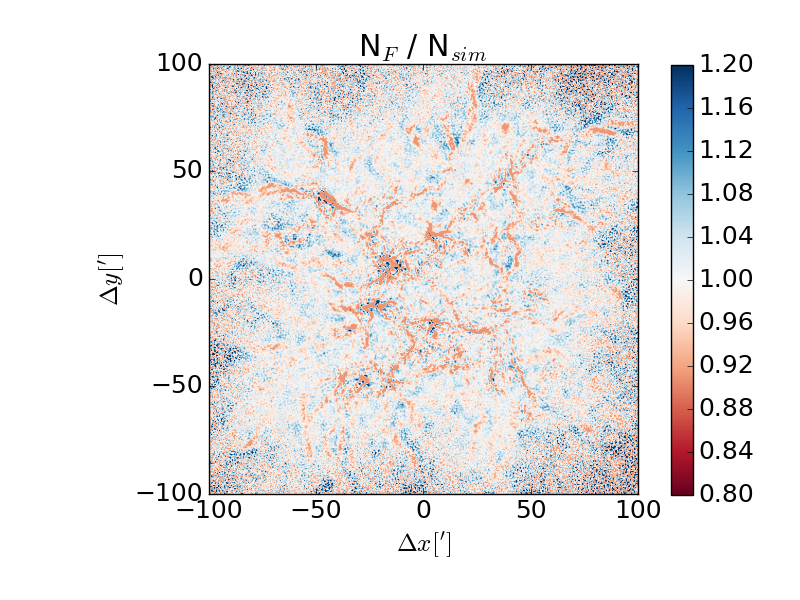}\\
\includegraphics[width=0.5\textwidth]{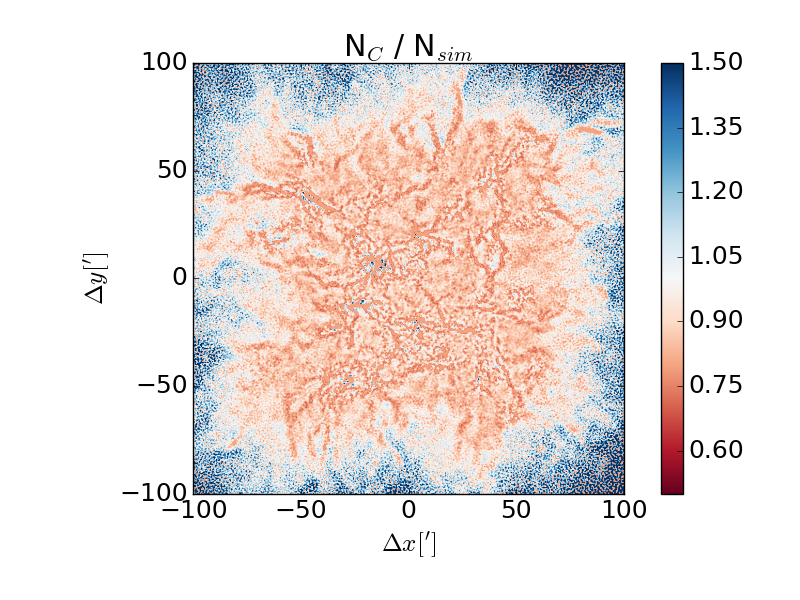}
\includegraphics[width=0.5\textwidth]{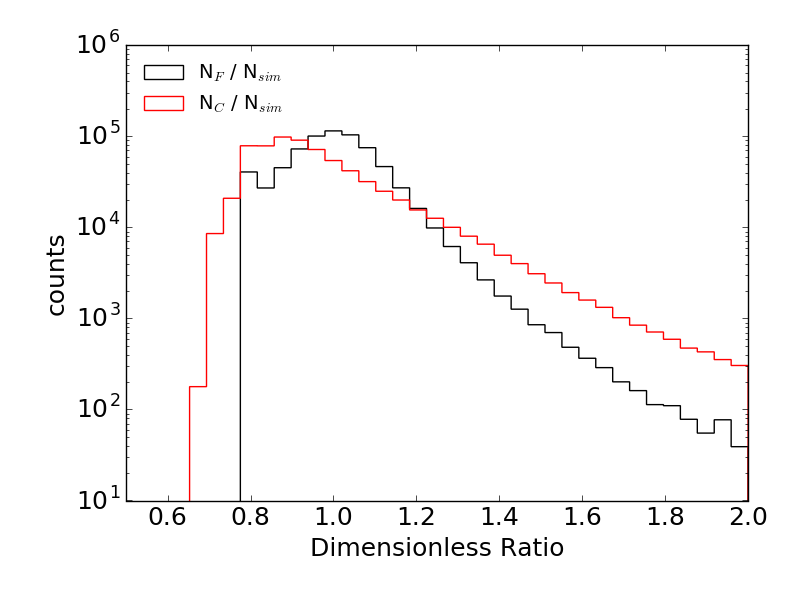}
\caption{\emph{Top left: }
		Column density map of the simulated molecular
		cloud in units of cm$^{-2}$.
		\emph{Top right: }Ratio between the feathered and 
		simulated column densities of the simulated molecular 
		cloud ($N_{F}$/$N_{sim}$). \emph{Bottom left: }
		Ratio between the constant--offset and 
		simulated column densities of the simulated molecular 
		cloud ($N_{C}$/$N_{sim}$).
		\emph{Bottom right: } Histograms of the 
		$N_{F}$/$N_{sim}$ (black) and $N_{C}$/$N_{sim}$ (red)
		distributions.}
  \label{fig:sim_cloud_N}
\end{figure*}

\begin{figure*}[t]
\includegraphics[width=0.5\textwidth]{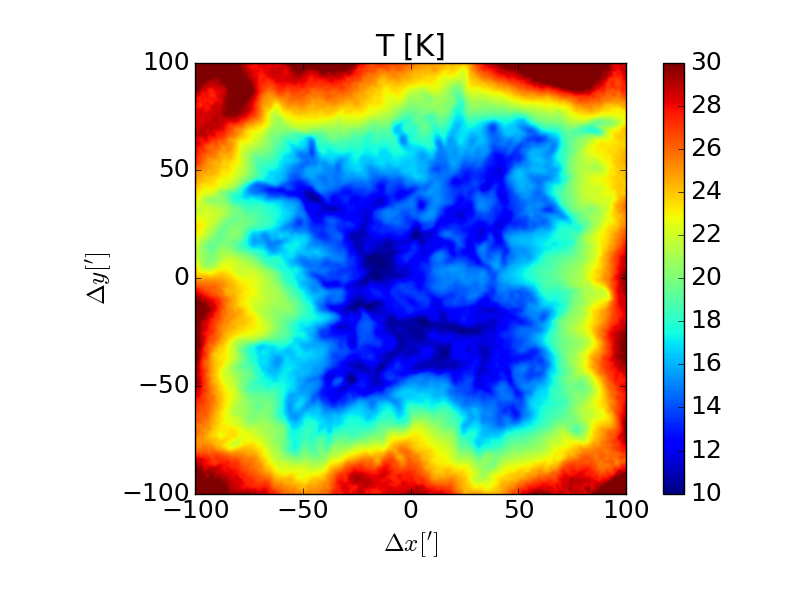}
\includegraphics[width=0.5\textwidth]{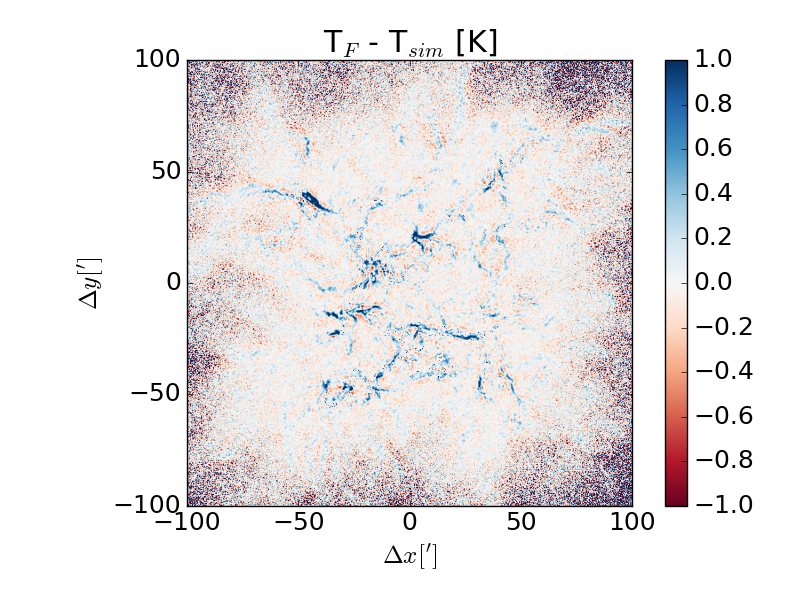}\\
\includegraphics[width=0.5\textwidth]{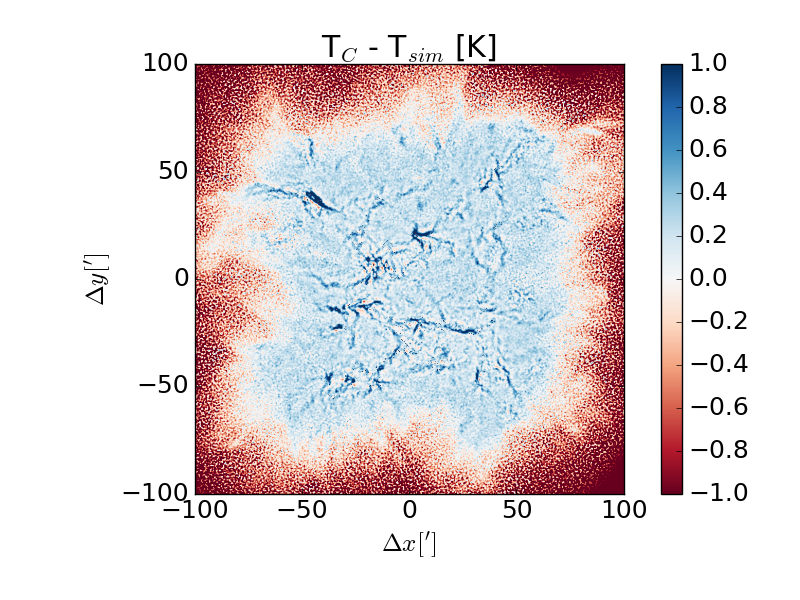}
\includegraphics[width=0.5\textwidth]{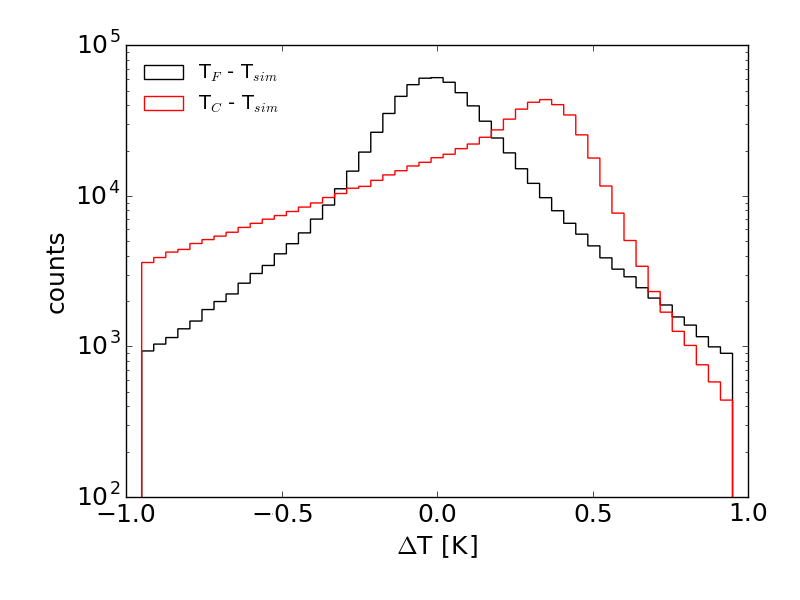}\\
\caption{Same as in Fig.~\ref{fig:sim_cloud_N} for the temperature.}
  \label{fig:sim_cloud_T}
\end{figure*}


We now test the performance of our method on a simulated $10^3$~\Msun\,
molecular cloud, for which the actual column density and temperature
distributions are known.  We used the Smooth Particle Hydrodyncamics
{\it SPH} code Gadget 2, with 24 million particles, starting from a
uniform density field and turbulent, random initial velocity
fluctuation with a {\it rms} mach number $\sim 8$.   For simplicity and speed of the calculations, our simulation was isothermal.  However, we assumed a polytropic  equation of state to produce a mean temperature map.  The internal temperature of the
cloud was, thus, given by
\begin{equation}
T = A n^\gamma ,
\end{equation}
where $n$ is density and 
with $A=215$ and $\gamma = -0.3 $ if $n \le 4.3\times10^5$ cm$^{-3}$,
and $A=5$ and $\gamma = 0.01$ if $n > 4.3\times10^5$ cm$^{-3}$.  Such
dependency is meant to be representative of the temperature of MCs,
which external parts are heated by the diffuse UV radiation of the
ISM, while their densest parts start becoming optically thick, the
cooling become less efficient, and thus the gas and dust grains become
coupled \citep[see, e.g.][]{jappsen05}.

To test our method, we need to construct column density 
and temperature maps with sizes of at least $2048^2$ pixels. 
For this purpose, we computed the total mass and the mean 
temperature of the {\it SPH} particles along each area element. 
Note that this was performed without taking into account 
the smoothing length in the simulations. Since 24 million 
particles distributed over 2048$^2$ pixels give a mean of 
$\sim$6 particles per beam, it is clear that, without considering 
the smoothing length, the map will contain a large amount 
of small structure noise. By not applying the smoothing 
length we verify that our algorithm is able to recover 
structure even in maps containing significant small-scale 
structure variation.
The resulting maps are displayed in the top
left panels of Fig.~\ref{fig:sim_cloud_N} and Fig.~\ref{fig:sim_cloud_T}.

\begin{figure}[t]
\resizebox{\hsize}{!}{\includegraphics{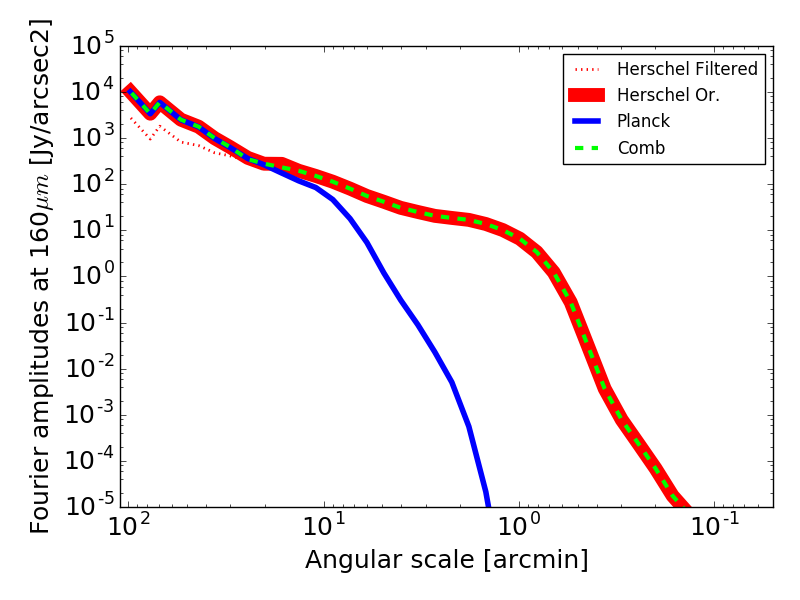}}\\
\caption{Radial averaged visibilities of the simulated 
	\planck (solid blue) data, simulated \herschel data 
	(solid red), filtered \herschel data (dotted red), and combined
	image generated by our method (dashed green). The scale is shown
	in units of arcminutes and the amplitudes of the visibilities in
	Jy/$\arcsec$ $^{2}$. The visibilities belong to simulated data  
	at 160\micron.}
  \label{fig:sim_radial}
\end{figure}

In order to test our method on the simulated clouds we
must generate the same data products retrieved from
the \herschel and \planck archives. Here we
explain the processing steps applied to obtain our 
\planck and \herschel simulated maps:

\begin{itemize}

\item[0.] \emph{Rotate the maps: } To obtain the most
		realistic possible test of our method we first 
		rotated the simulated molecular cloud maps 45 
		degrees. With this step we test possible 
		effects introduced when we crop and rotate the
		real maps in the first step of our method.
		
\item[1.] \emph{Obtaining \planck and \herschel emission 
		maps}: With our simulated maps and, for simplicity,
		 assuming a constant dust spectral index, $\beta$=1.8,
		consistent with~\citet{oh94}, we use 
		Eq.~\ref{eq:sed} and Eq.~\ref{eq:int-sed} to
		obtain the ideal observations of our simulated
		clouds for \herschel and \planck, each at their 
		own wavelengths (Sect.~\ref{data:herschel} and
		Sect.~\ref{data:planck}). 
		
\item[2.] \emph{Adding realistic noise to the 
		emission maps:} The emission maps obtained in
		step 1 are highly idealized. We therefore 
		add realistic noise to our ideal \planck and
		\herschel emission maps to test our method
		under more realistic conditions. We estimate the
		noise of \planck and \herschel in the Fourier 
		space using the actual HiGal--11 maps for this purpose.
		We assume that the Fourier amplitudes of \planck
		and \herschel have two main components: the signal, encoded in the 
		Fourier amplitudes at any given scale, and 
		the noise, which causes scatter in the signal at any given scale.   
		This noise includes observational and ``artificial''
		effects (e.g., gradients) created by image processing
		pipelines. We use the \emph{rms} of the Fourier amplitudes 
		to estimate the \herschel and \planck noise. We then convolve this 
		noise with our ideal emission maps. With this method,
		we include in our simulation artificial effects in \herschel
		and \planck data (e.g., saturation of the IRAS data in
		the \planck dust model, artificial gradients in \herschel).

\item[3.] \emph{Obtaining more realistic data products 
		from the \planck all-sky model of dust emission:} 
		As explained in Sect.~\ref{data:planck}, the \planck
		all sky model of dust emission provides three
		data products: temperature, optical depth, and 
		dust spectral index. To test our model we 
		used the realistic \planck emission maps obtained in
		step 2 to generate our temperature, optical depth, and 
		dust spectral index datasets. We use the
		same procedure followed in~\citet{planckXI14}.
		The \planck datasets obtained in this step will be
		the starting point to apply our method to the simulations.
				
\item[4.] \emph{Filtering of the \herschel maps: } The goal
		of our method is to correct the background emission
		of the \herschel maps applying multi-scale corrections
		derived from \planck. To test how our method recovers 
		possible large scale variations measured by \planck,
		we artificially filter the \herschel maps at large scales.
		We illustrate this procedure, for the case
		of 160\micron, in Fig.~\ref{fig:sim_radial}.
		This image shows the radial averaged Fourier amplitudes of
		our \planck and \herschel simulated maps (blue and red
		solid lines respectively). We then filter the \herschel
		data at scales larger than 30$\arcmin$, resulting in the
		red dotted line shown in Fig.~\ref{fig:sim_radial}.
		The green dashed line in Fig.~\ref{fig:sim_radial} shows
		the radial averaged Fourier amplitudes of the 
		feathered image resulting of  applying our method to 
		the simulated datasets. Note that it follows nicely the
		red solid line of the original \herschel data, showing that
		our method achieves the goal of recovering filtered emission
		in \herschel. 
		
\end{itemize}

We have now obtained the entire dataset needed to 
apply our method as it has been done in previous
sections to the real data. We run our method
as explained in Sect.~\ref{sec:method} on the
\herschel and \planck simulated datasets obtained
in steps 4 and 3 respectively. As in previous sections,
the application of our method generates constant--offset
(the \herschel maps corrected only with the zeroth
Fourier order) and feathered flux maps that we further 
process to obtain their column densities and 
temperatures following App.~\ref{sec:get-nh-t}. 

In the top left panel of Fig.~\ref{fig:sim_cloud_N} we show 
the column density map of our simulated
molecular cloud. 
In the top right panel of Fig.~\ref{fig:sim_cloud_N} we compare our 
feathered column density distribution with that of the simulated cloud.
The feathered and 
simulated column densities agree within 10\% in general at large scales.  
In the densest regions ($N>10^{23}$ cm$^{-2}$) of the 
simulated cloud our method tends to systematically 
underestimate the column densities by values up to 15--20\%. 
This effect is mainly caused by the simplifying assumption
of a constant dust spectral index introduced in the 
step 1. We also note the absence of edge
effects in our feathered maps, which demonstrates that cropping and rotating
the maps do not generate artifacts in our maps. 
In the bottom left panel of Fig.~\ref{fig:sim_cloud_N} we show the 
comparison between the constant--offset and simulated column densities.
These have some features in common with the behavior outlined
above for the feathered case: systematically underestimated column densities
in the densest parts of the cloud. Most importantly, the constant--offset
maps show a bias towards column densities 10\% lower than those
of the simulated cloud. This is better seen in the histogram 
at bottom right panel of Fig.~\ref{fig:sim_cloud_N}, that peaks
at $N_{C}/N_{sim}\approx0.90$.
In contrast, the $N_{F}/N_{sim}$ histogram peaks at 1.
Furthermore, the constant offset method is also more prone to 
overestimate column densities (specially in the low
column density areas of the cloud) than our feathered maps, as shown by the
wider distribution of its histogram, compared to that of our feathered maps.

In Fig.~\ref{fig:sim_cloud_T} we compare the temperature results. 
In the top left panel of Fig.~\ref{fig:sim_cloud_T} we show the temperature
distribution of the simulated cloud, which is compared to the 
feathered and constant--offset temperatures in the top right and
bottom left panels respectively. 
As in the case of column densities,
there is very good agreement (within half a K)
between the simulated and feathered temperatures. The feathered 
temperatures tend to be higher (up to 1 K) than the simulated 
temperatures in the coldest regions of molecular clouds. 
This effect, also seen in the constant-offset temperatures, 
is likely connected to the assumption of a constant dust 
spectral index done in the step 1, and is also reflected in 
the column density maps (see above). 
The histogram of Fig.~\ref{fig:sim_cloud_T} shows that the 
constant--offset temperatures are biased towards 
higher values of 0.35 K, while the distribution of 
feathered temperatures peaks at $\Delta$T = 0 K, indicating
no bias. Both temperature distributions have, however, large 
widths (up to $\pm$1 K), with the feathered distribution 
being highly symmetrical, in contrast to the 
constant--offset distribution.

These results
show that our method recovers better the original N$_{\rm H}$ and T
information, thanks to the combination of large and small scale
information.  Furthermore, our method accomplishes this without
introducing additional artifacts due to the data treatment.

\section{Conclusions}\label{sec:conc}

The goal of this paper is to derive \planck--based multi--scale
corrections for the \herschel images at 160~\micron, 250~\micron,
350~\micron, and 500~\micron. We achieve this goal by linearly
combining \herschel and \planck data in Fourier space. We test our
method in two different star forming molecular cloud regions.  We
further processed the feathered maps to obtain the column densities
and temperatures of the regions studied. We compare our feathered
column densities and temperatures with those obtained with previous
methods: corrected the flux scale of \herschel images adding a
constant--offset value obtained from comparisons with \planck data. We
finally demonstrate the performance of our method in simulations.
Here we summarize the main results of our paper.

\begin{itemize}

\item We combine the \planck and \herschel datasets at an effective
  scale $\kappa_{eff}$. This effective scale is calculated separately
  for each region as the angular scale at which the residuals between
  the combined image visibilities and the original \herschel and
  \planck visibilities are minimum.  The effective scale has values
  between the \planck resolution ($5\arcmin$) and
  $\kappa_{eff}=\infty$. The latter is equivalent to applying only the
  Fourier zeroth mode correction to \herschel, and is mathematically
  identical to the previously used constant--offset correction. Our
  method is therefore a general method to correct the \herschel flux
  scales with the constant--offset correction arising naturally as a
  special case.
		
\item Our method can be generically applied to any combination of
  image estimators containing different angular resolutions.

\item In the HiGal--11 field, our feathered column densities exhibit
  higher (lower) N$_{\rm H}$ values in (out of) the Galactic plane region,
  compared to the constant--offset method.  In general, a similar
  effect is seen in Perseus in the areas surrounding NGC 1333, which
  also exhibits higher N$_{\rm H}$ values compared to previous methods.
  We show that N$_{\rm H}$ values calculated based on the constant--offset
  method can be discrepant by factors of $\sim50$\% or more, but
  typically span variations of $\sim30$\% over significant portions of
  the images.

\item In the two regions shown in this paper our feathered column
  densities recover more low column material, and the discrepancies
  with the previous method are most significant at the lower end of
  the column density distribution, near N$_{\rm H}$
  $\,\sim\,10^{22}$~cm$^{-2}$.  Above this value, we find generally
  acceptable agreement with previous methods. As most molecular cloud
  mass resides at low N$_{\rm H}$ values, a proper treatment of the column
  densities and temperatures is needed to better constrain fundamental
  physical parameters such as the gravitational potential.

\item We also apply our method to simulated molecular cloud data,
  where the actual temperature and column density distributions are
  known. We simulated the \planck and \herschel observations for the
  simulated cloud, including noise.  We then applied artificial
  filtering to the simulated \herschel data.  The comparison between
  the input and output N$_{\rm H}$ and T maps reveals that our method
  successfully recovers the emission filtered out from the \herschel
  data.  Therefore we conclude that our method is accurate and does a
  better job at reconstructing the missing background emission than a
  constant--offset correction alone would.

\end{itemize}

We make the data in this paper publicly available. Furthermore, 
our technique can be applied to the entire \herschel
science archive. This is the goal of a follow up paper.

\begin{acknowledgements}
  JA and AS thank the referee for a constructive report whose line of
  inquiry led to significant improvements to this work.  JA and AS
  thank Andrew Gould for insightful technical discussions that
  resulted in significant improvements to the method.  The work of JA
  is supported by the Sonderforschungsbereich (SFB) 881
  \textquotedblleft The Milky Way System \textquotedblright and the
  International Max-Planck Research School (IMPRS) at Heidelberg
  University. AS is thankful for funding from the ``Concurso Proyectos
  Internacionales de Investigaci\'{o}n, Convocatoria 2015'' 
  (project code PII20150171) and the BASAL Centro de Astrof\'{i}sica 
  y Tecnolog\'{i}as Afines (CATA) PFB-06/2007. 
  J.B.-P. acknowledges UNAM-PAPIIT grant number IN110816, 
  and to UNAM’s DGAPA-PASPA Sabbatical program. He also is 
  indebted to the Alexander von Humboldt Stiftung for its invaluable support.
  This paper includes data from \herschel , a European
  Space Agency (ESA) space observatory with science instruments
  provided by European--led consortia and with important participation
  from NASA.  This papers makes use of data provided by \planck , a
  project of the European Space Agency (ESA) with instruments provided
  by the o scientific consortia funded by ESA member states and led by
  Principal Investigators from France and Italy, telescope reflectors
  provided through a collaboration between ESA and a scientific
  consortium led and funded by Denmark, and additional contributions
  from NASA (USA).  This research made use of Montage. It is funded by
  the National Science Foundation under Grant Number ACI-1440620, and
  was previously funded by the National Aeronautics and Space
  Administration's Earth Science Technology Office, Computation
  Technologies Project, under Cooperative Agreement Number NCC5-626
  between NASA and the California Institute of Technology.
\end{acknowledgements}

\bibliographystyle{aa} 
\bibliography{bibliography}

\appendix

\section{Our interpolation functions}\label{sec:weights}

As shown in Eq.~\ref{eq:fft-comb}, the combination of
\planck and \herschel datasets in the Fourier space
includes the definition of two $uv$--scale dependent 
interpolation functions, $w_{P}(\kappa)$ and $w_{H}(\kappa)$
respectively. In the text we explain that the canonical
feathering technique defines both functions as Gaussian,
based on the approximation of telescope beam profiles as
Gaussian functions. However,~\citet{csengeri16} have shown that
alternative interpolation functions can be successfully used
to combine the data of two single dish telescopes in the
Fourier space.

\begin{figure}[t]
\resizebox{\hsize}{!}{\includegraphics{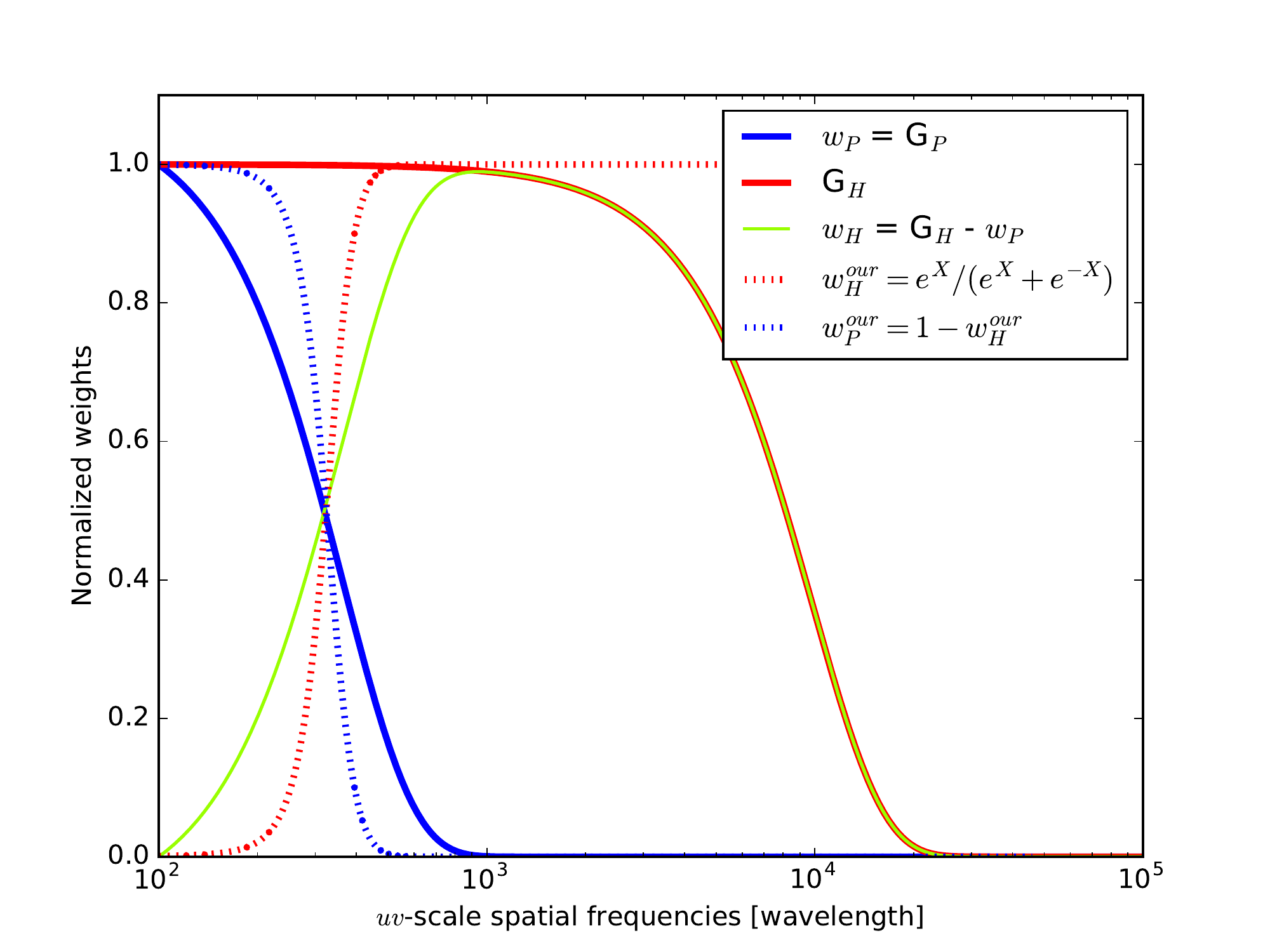}}
\caption{Fourier scale dependent functions 
	applied in the classical feathering 
	technique~\citep{stanimirovic02}. The blue (red)
	solid line shows the \planck (\herschel at 160\micron) beam 
	approached by a Gaussian of 5$\arcmin$ (12$\arcsec$) 
	in the Fourier domain. The green line shows the scale--dependent
	functions applied to the \herschel data following the classical
	feathering algorithm (see Sect.~\ref{sec:comb}). The
	dotted red (blue) line shows the interpolationg functions
	used in this paper for \herschel, $w_{H}^{our}$ (\planck,
	$w_{P}^{our}$) assuming $\kappa_{eff}=5\arcmin$. 
	The functional form of these interpolation functions is
	shown in Eq.~\ref{eq:wplanck} (Eq.~\ref{eq:wpacs}).}
  \label{fig:uv-scheme-app}
\end{figure}

Specifically for our method, the interpolation functions must
fulfill one condition: they must transition smoothly from 
the \herschel regime at small \textit{uv}--scales to
the \planck regime at large \textit{uv}--scales with
no loss in flux. To this aim, we decided to  
define $w_{H}$ and $w_{P}$ as:

\begin{equation}\label{eq:wplanck}
w_{H}(\kappa) = \frac{e^{x}}{e^{x}+e^{-x}};\\ x = Q\left( \frac{\kappa}{\kappa_{eff}}-1\right) ,
\end{equation}

where $Q$ is a factor defining the steepness
of the 
interpolation functions in the transition, and $\kappa_{eff}$ 
is the \emph{effective} scale at which we combine
the \herschel and \planck datasets.
Note that Gaussian functions have no equivalent property
to $Q$. 
Our interest on controlling the steepness of the
interpolation functions responds to the condition of losing
no (or negligible) flux when transitioning from 
\planck to \herschel scales. 
The interpolation functions to be applied to the \planck data are defined as
\begin{equation}\label{eq:wpacs}
w_{P}(\kappa) = 1 - w_{H}(\kappa),
\end{equation}
filling the requirement $w_{P}+w_{H}=1$ for every scale
(see Fig~\ref{fig:uv-scheme-app}).
Note that in the standard feathering the weights 
sum to the Gaussian beam of the interferometer. The reason 
is reducing possible noise at the smallest $uv$--scales.
However, we want to keep all power of \herschel at
small scales. We therefore require that the sum of our interpolation functions is
one at all scales. In Fig.~\ref{fig:uv-scheme-app} we show
the comparison between the canonical Gaussian weights used in
the feathering for interferometry and those used in our paper.
We validate our non standard interpolation functions in the 
simulations show in Sect.~\ref{sec:sims} showing that we 
reproduced properly the distributions.

\section{N$_{\rm H}$ and temperature fitting: modified black-body fitting}\label{sec:get-nh-t}

We provide a brief summary here and refer the reader to~\citet{stutz15}
and~\citet{stutz10} for further details.

We convolve the feathered data to the beam of \herschel
500$\,\mu$m (FWHM $\sim\,$36\arcsec) using convolution 
kernels from \citet{aniano11}. 
We then re--grid the data to a common coordinate system,
using an 14\arcsec pixel scale.  With the surface densities
of the four wavelengths we obtain an SED for each pixel.
We fit each pixel SED using an MBB function: 
\begin{equation}
  S_{\nu} = \Omega\,B_{\nu}(\nu,T_{\rm d})\,(1-e^{-\tau(\nu)}),
\end{equation}
where $\Omega$\ is the beam solid angle, $B_{\nu}(T_{\rm d})$ 
is the \planck function at a dust temperature $T_{\rm d}$, 
and $\tau(\nu)$\ is the optical depth at frequency $\nu$.
We define the optical depth as $\tau(\nu) = N_{\rm H}\,m_{\rm
  H}\,R_{gd}^{-1}\,\kappa(\nu)$, where $N_{\rm H} = 2\,\times\,N({\rm
  H_2}) + N({\rm H})$ is the total hydrogen column density, $m_{\rm
  H}$ the mass of the hydrogen atom, $\kappa_{\nu}$ the dust opacity, 
  and $R_{gd}$ the gas--to--dust ratio,
assumed to be 110~\citep{sodroski97}.  We use the dust opacities
listed in the column 5 in Table~1 of \citet{oh94}: 
dust grains with thin ice mantles after $10^5$~years of
coagulation time at an assumed gas density of $10^6$~cm$^{-3}$. 
The systematic effects introduced when assuming a different dust model
are discussed in~\citet{stutz13} and~\citet{laun13}.
The choice of dust model, along with the adopted $R_{gd}$
value, likely dominate the systematic uncertainties.

We use a two-step method for applying the color and beam size
corrections to the pixel SEDs. We fit the uncorrected fluxes to obtain
a first estimate of the temperature. We then use this temperature to
apply the corrections as described in the SPIRE and PACS instrument
handbooks.  We then repeat the fit to the corrected SED.

\end{document}